\documentclass[11pt]{article}
\usepackage{graphicx}

\usepackage{amssymb}
\usepackage{amsmath}
\usepackage{amsfonts}

\newtheorem{theorem}{Theorem}[section]

\newtheorem{definition}[theorem]{Definition}
\newtheorem{lemma}[theorem]{Lemma}
\newtheorem{remark}[theorem]{Remark}

\newcommand{\pf}{{\noindent {\bf Proof: }}}

\newcommand{\qed}{$\ \Box $\medskip }

\newcommand{\W}{{\cal W\/}}
\newcommand{\T}{{\cal T}}
\newcommand{\B}{{\cal B}}
\newcommand{\M}{{\cal C}}
\newcommand{\D}{{\cal D}}
\newcommand{\PR}{{\cal P}} 

\title{Ranking pages and the topology of the web}

\author{Argimiro  Arratia\footnote{Research partially supported  by BASMATI MICINN project (TIN2011-27479-C04-03) and by
SGR2009-1428 (LARCA) 
 and SINGACOM (MTM2007-64007)}  \\ 
Departament de Llenguatges i Sistemes Inform\`atics,\\
Universitat Polit\`ecnica de Catalunya, 
Barcelona, Spain\\
argimiro@lsi.upc.edu
\and 
Carlos Mariju\'an
\footnote{Supported   
by Spanish Government MICINN  project SINGACOM (MTM2007-64007)} \\
Departamento de Matem\'atica Aplicada\\  
Universidad de Valladolid, Valladolid, Spain,\\
marijuan@mat.uva.es
}

\date{}

\begin{document}

\maketitle

\begin{abstract} 
This paper presents our studies on the  rearrangement of links in the structure
of websites for the purpose of improving the valuation of a  page or group of pages as
established by a ranking function as Google's PageRank.
We build our topological taxonomy starting from unidirectional and bidirectional rooted trees, and  up to more complex hierarchical structures as cyclical rooted trees (obtained by closing cycles on bidirectional trees) and PR--digraph  rooted trees (digraphs whose condensation digraph is a rooted tree that behave like cyclical rooted trees). 
We give different modifications on the structure of these trees and its effect 
on the valuation given by the PageRank function. 
We derive closed formulas for the PageRank of the root of various types of trees, and establish a hierarchy of these topologies in terms of PageRank.


\end{abstract}

\section{Introduction}\label{sec_Intro}
Google is still today's most popular search engine for the World Wide Web, 
and the key to its success has been its PageRank algorithm \cite{BP98}, which ranks  documents based primarily  on  
the link structure of the web.
Simply put, PageRank considers
a link from a page $H$ to another page $J$  
as a weighted vote from $H$ in favour of  the importance of $J$, 
where the weight of the vote of $H$ is itself determined by the number of links
(or voters) to $H$. 
Therefore, part of the game of the electronic business today is to find ways of lifting   a page's link popularity, and specifically the PageRank, 
by either obtaining the vote of a very important page (which is unlikely) or 
manufacturing a large set of pages that would be ``willing'' to link to a client's page. For the latter solution, known as {\em link farms} in the jargon of the Search Engine Optimisation (SEO) community, much care must be taken
since it is widely believed that Google had tuned up its original PageRank algorithm to detect fictitious  linking and similar forms of spamming (e.g. the 2003 ``Florida'' update, see \cite{florida}).

At the heart of the challenge of improving a page PageRank value is 
the role played by the topology of the web. 
This is a widely recognised fact as
there can be found in the internet many SEO analyses of link patterns, 
together with tips on how to rearrange these to raise the PageRank of specific pages. 
On the theoretical side, having acknowledged that the World Wide Web should be treated as a directed graph, there are various publications that propose different graph decompositions on regular patterns, as a way to improve  PageRank computation  (e.g. \cite{Arasu02}, \cite{Kamvar}), and news that suggests that newly acquired  technology by Google, in the hope to enhance PageRank, is based on localisation of the computations on certain tree structures underlying the Web (\cite{googlepress}, \cite{Olsen03}).

Motivated by these graph combinatoric challenges particular to the Web, 
we have studied the PageRank formula from a mathematical perspective, and
its relation with the
web site's topology,  
with the 
twofold goal of accelerating the computation of Page\-Rank and maximising its value for an specific page or set of pages. 
We summarise here all our findings starting from (unidirectional) rooted trees and up to more complex hierarchical structures. 
Ultimately, 
our academic goals are to disclose some of the graph combinatorics underlying the World Wide Web and this popular ranking function, and
to contribute to the   mathematical foundation of many 
heuristics and ad hoc rules in used by the SEO community in its attempt to 
tweak the valuations assigned by PageRank.

\section{Some preliminaries on Graph Theory}
In this paper we will use some standard  concepts and results about directed graphs, 
which we detail in this section in order to fix our notation.

By a {\bf digraph} $\D$ we mean a pair $\D=(V,A)$ where $V$ is a finite nonempty set and 
$A\subset V\times V \setminus \{(v,v): v\in V\}$. Elements in $V$ and $A$ are called {\bf vertices} and {\bf arcs} 
respectively. For an arc $(u,v)$ we will say that $u$ is {\bf adjacent to} $v$, and 
we may sometimes also use $uv$ to denote an arc $(u,v)$.
 The {\bf order} and the {\bf size} of $\D$ are, respectively, $Card(V)$
 and $Card(A)$. If $v$ is a vertex, the {\bf in-degree}, $id(v)$, of $v$ is the number of arcs $(u,v)$ in $A$.
 Similarly, the {\bf out-degree}, $od(v)$, of $v$ is the number of arcs $(v,u)$ in $A$.

A sequence of vertices $v_1v_2\dots v_q, q\geq 2,$ such that $(v_i,v_{i+1})\in A$ for
$i=1,2,\dots ,q-1$ is a {\bf walk} of {\bf length} $q-1$ {\bf joining} $v_1$ {\bf with} $v_q$ 
or more simply a $v_1$--$v_q$ {\bf walk}. If the vertices of $v_1v_2\dots v_q$ are distinct the walk is called a 
{\bf path}.
A {\bf cycle of length} $q$ or a $q$-{\bf cycle} is a path  $v_1v_2\dots v_q$
closed by the arc $v_qv_1$.
A digraph is {\bf acyclic} if it has no cycle. 
By a 
{\bf semipath joining } $v_1$ {\bf with} $v_q$  we mean a sequence of distinct vertices $v_1v_2\dots v_q, q\geq 2,$ such that 
 $(v_i,v_{i+1})\in A$ or $(v_{i+1},v_i)\in A$ for $i=1,2,\dots ,q-1$.

A digraph is {\bf connected} if for each pair $u$ and $v$ of distinct vertices, there is a semipath joining $u$
 with $v$.
By a {\bf subdigraph} of the digraph $(V, A)$ we mean a digraph $(W,B)$ such that $W\subset V$ and $B\subset A$.
 The subdigraph is called a {\bf partial digraph} when $W=V$. The {\bf induced subdigraph} by the digraph $(V,A)$ 
on $W\subset V$ is the digraph $(W,A/W)$ where $A/W=A\cap (W\times W)$.

For an acyclic digraph there exists at least one vertex $v$ (resp. $u$) such that $od(v)=0$ (resp. $id(u)=0$). 
Such vertex will be called a {\bf maximal} (resp. {\bf minimal}) in the digraph. Moreover, the vertices in an 
acyclic digraph $(V, A)$ can be distributed by {\bf  levels} $N_0, N_1, \dots $,
where
$N_0=\{ v\in V: v \; \mbox{is maximal in} \; (V, A)\}\ $
 and, recursively for $p>0$,
$$N_p=\{v\in V\setminus\bigcup_{i=0}^{p-1}N_i: v \; \mbox{is maximal in the induced subdigraph on} \;
  V\setminus \bigcup_{i=0}^{p-1}N_i \}$$

Thus one has a partition of $V$, $V=N_0\cup N_1\cup \dots \cup N_h$, $h$ being the {\bf height of the digraph}, i.e. the last 
index such that $N_h \neq \emptyset $.

\section{Short Introduction on PageRank}\label{introPR}
 
The mathematical view of the World Wide Web is as a digraph $\W = (V,A)$, where 
a vertex represents   any document posted on the web (a {\em page}), and  an 
arc $(b,a)$ indicates that there is a  link from page $b$ to page $a$. In this setting, Brin and Page proposed  
in \cite{BP98} to evaluate each page in the Web with a positive real number, which they named its PageRank, given by the formula (in its refined version from \cite{BMPW98}):
 
\begin{equation}\label{prfmla1}
\PR(a) = \frac{1 - \alpha}{N}  + \alpha\sum_{(b,a) \in A} \frac{\PR(b)}{od(b)} 
\end{equation}
where $\PR(a)$ is the PageRank of page $a$, $od(b)$ is the number of links going out of page $b$ (the out-degree of $b$), $\alpha$ is a constant that can take any real value in the interval $(0,1)$ (although  Brin and Page always prefer to  set it to $0.85$), $N$ is the total number of pages of the Web, and the sum is taken over all pages $b$ that have a link to $a$. 
The motivation, given by the authors, is that  formula (\ref{prfmla1}) models the behaviour of a random surfer of the Web who,  being at a certain page $b$, either follows one of the links shown in that page with probability $\alpha$, or jumps to any other page with probability $1-\alpha$, disregarding the contents  of the pages. The probability of choosing a link in $b$ that takes him to page $a$ depends on the number $od(b)$ of links out of $b$; so $\PR(b)/od(b)$ is the contribution of $b$ to the PageRank of $a$ amortised by $\alpha$.
In this setting, the PageRank of $a$ is the probability of a user reaching 
page $a$ directly or after following all appropriate links, and   the sum of the PageRank  of all the pages is 1, and so, forms a probability distribution over the Web
(see \cite{BGS05} and \cite{LM04}).

Yet another view of PageRank
is the analytical  formulation  given by
 Brinkmeier (see \cite{Brink06}), who conceived  the PageRank function as a power series. In this setting,  a formula is given that highlights the fact that the ranking of a vertex $v$, as assigned by PageRank,  depends on the weighted contributions of each vertex in  every walk that leads into $v$,
 being these contributions higher in value for vertices that are  nearer in distance from $v$.  

For a given walk $\rho = v_1 v_2 \ldots v_n$ in the graph $(V,A)$, define the {\bf branching factor} of $\rho$ by the formula
$$D(\rho) = \frac{1}{od(v_1)od(v_2) \cdots od(v_{n-1})}$$
Then, for any vertex $a \in V$, we have
\begin{equation}\label{brinPR}
{\PR}(a) = \frac{1-\alpha}{N}\sum_{w \in V}
 \sum_{\rho\,:\,  w\, \stackrel{*}{\longrightarrow}\, a} \alpha^{l(\rho)}D(\rho) 
\end{equation}
where $\rho: w \stackrel{*}{\longrightarrow} a$ denotes a walk $\rho$ starting at vertex $w$ and ending in vertex $a$, and $l(\rho)$ is the length of this walk $\rho$.

\section{Ranking vertices on  trees}\label{PR4tree} 

Our starting case study is the set of {\em rooted   trees}, 
where a {\bf tree with root} is an acyclic digraph for which there exists a maximal
vertex $r$ (the root), such that for every vertex $v \not= r$ there is a unique
$v$--$r$ path. We denote a tree with root $r$ as $\T^r$. Thus, a tree $\T^r$ is a connected graph, its root $r$ is unique and all
  vertices distinct from $r$ have
out--degree  1, whilst  the in--degree may vary.  
Vertices with in--degree $0$ are called {\bf leaves}.
The root is the targeted page for improving its PageRank valuation.  
The {\bf height} of a vertex in a rooted tree   is the length of   the path from the vertex to the root. The { level} $k$ of a rooted tree is the set of vertices with height $k$;  the root is at level $N_0$. The {\bf height} of a rooted tree is the
length of the longest  path from a leaf to the root. 

\begin{remark}\label{treeloco}  
Since we are interested in studying the behaviour of 
PageRank
 when localised in certain subdigraphs of the Web digraph, we think, in particular, of our trees as local closed web sites. This 
means that
the  value of $N$ in formula 
(\ref{prfmla1}) is the number of vertices in the tree.
\qed
\end{remark}
  
Our first result shows that to compute the 
PageRank   
of the root of a tree all we need to do is count the number of vertices at each level of the tree.
\begin{theorem}\label{pr4treethm}
If a   rooted tree has $N$ vertices and height $h$, then the PageRank of its root $r$ is given by the formula
\begin{equation}\label{pr4tree}
\PR(r) = \frac{1 - \alpha}{N}%
\sum_{k = 0}^{h} \alpha^{k} n_k
\end{equation}
where $n_k := |N_k|$ is the number of vertices of the $k$th--level, $N_k$, of the tree. 
\end{theorem}
\pf
Below we use $b \in N_{k} : b \to a$ to indicate that vertex $b$ at level $N_k$ has a link to $a$. 
Assume the first level of the tree $N_1 = \{a_1, \ldots, a_{n_{_1}}\}$. Then, according to equation (\ref{prfmla1})
\begin{eqnarray*}
\PR(r) &=& \frac{1 - \alpha}{N} + \alpha\sum_{a \in N_{1}} \PR(a)
\ =\ \frac{1 - \alpha}{N} + \alpha  \left(
\left(\frac{1 - \alpha}{N} + \alpha\sum_{b \in N_{2} : b \to a_1} \PR(b) \right) \right.  + \\
& & \quad \ldots \quad +\
\left. 
\left(\frac{1 - \alpha}{N} + \alpha\sum_{b \in N_{2} : b \to 
a_{n_{_1}}} \PR(b) \right) \right) 
\end{eqnarray*}
 The index sets $\{b \in N_{2} : b \to a_i\}$, for $i = 1, \ldots, n_1$,  are pairwise disjoint; therefore,
\begin{eqnarray*}
\PR(r)  
&=& \frac{1 - \alpha}{N}(1 + \alpha n_{_1}) +  \alpha^2\sum_{b \in N_{2}} \PR(b)
\end{eqnarray*}

Repeating  the above manipulations on levels $N_2$, $N_3$, and 
up to level $N_{h-1}$, we have
$$\PR(r) = \frac{1 - \alpha}{N} 
\sum_{k=0}^{h-1} \alpha^k n_{_k} +
\alpha^{h} \sum_{b \in N_{h} } \PR(b)$$ 
At the last level $N_h$ all vertices are leaves, which have no in-coming arcs, hence the 
PageRank 
of any $b \in N_h$ is $(1 - \alpha)/N$. Then
$$\alpha^h\sum_{b \in N_{h} } \PR(b) =\frac{1 - \alpha}{N} \alpha^h n_{_h}$$
and the result follows.
\qed

\begin{remark} 
 Theorem \ref{pr4treethm} shows that we can do any rearrangements of links between two consecutive levels of a web  set up as a rooted tree, and the Page\-Rank of the root will be  the same.
\qed
\end{remark}

\begin{remark}\label{uniTreeDef}
Due to Theorem \ref{pr4treethm}, we will from now on describe 
a rooted 
tree $\T^r$, with root $r$ and
$h\ge 0$ levels,
each of cardinality $n_0 = 1$, $n_1$, \ldots, $n_h$,   
as the string
$\quad \T^r = 1n_1 \ldots n_h$.  Also the PageRank for the root
$r$ of $\T^r$, or for any other vertex seemed as  the root of a subtree in $\T^r$, 
will depend on the height and 
the number of vertices at each level of $\T^r$. Henceforth,  we write PageRank of $r$ in the tree $\T^r$ as a function of the height $h$, and denote it $\PR(h)$. \qed
\end{remark}

For some regular topologies we can have nice closed formulas for their PageRank. Some examples follow below.

 \subsection{$m$-ary trees}\label{narytree}
For $m,h \ge 1$ , let $\T^r_m(h)$ be the full {\bf $m$--ary tree of height}  $h$, 
i.e. a tree of height $h$ whose vertices, except by the leaves, have in--degree $m$.
 The $1$--ary tree of height $h$, $\T^r_1(h)$, is a path of length    $h$. 
For $m > 1$, $\T^r_m(h)$ has $m^k$ vertices at each level  $k = 0, 1, \ldots, h$, and the total number of vertices is $(m^{h+1} -1)/(m-1)$. Using Theorem \ref{pr4treethm} we can quickly calculate the PageRank for the root $r$ (which depends on the height $h$ and fixed arity $m$, and so we denote $\PR_{m}(h)$). This is
$$\PR_{m}(h) = (1-\alpha)\frac{m-1}{m^{h+1} - 1}\sum_{k=0}^h m^k  \alpha^k = 
(1-\alpha)\left( \frac{m-1}{m^{h+1} - 1}\right)\frac{(m\alpha)^{h+1} - 1}{m\alpha -1}$$
and for the $1$--ary tree
$$\PR_{1}(h) = \frac{1-\alpha^{h+1}}{h+1}$$

\subsection{Binomial trees}\label{binotree}
A binomial tree is at the core of fundamental data structures such as heaps, and hence, it qualifies as a good candidate for a website's topology.\footnote{Goodness as always is understood in terms of PageRank.} 

We use $\T^r_b(h)$ to denote the full {\bf binomial tree of height} $h$. 
We recall from \cite[\S 9.1]{CLR} that $\T^r_b(0)$ consists of only one vertex --the root-- and, inductively, $\T^r_b({h+1})$ is two copies of 
$\T^r_b(h)$ joint with an arc from the root of one of the $\T^r_b(h)$ to the root of the other.
At each level $k = 0, 1, \ldots, h$, $\T^r_b(h)$ has ${h \choose k}$ vertices, and the total number of vertices in $\T^r_b(h)$ is 
$\sum_{k=0}^h {h \choose k} = 2^{h}$.
Using Theorem \ref{pr4treethm} we get a nice formula to easily calculate the PageRank for the root $r$ of $\T^r_b(h)$, namely
$$\PR_{b}(h) = \frac{1-\alpha}{2^h} \sum_{k=0}^h {h \choose k} \alpha^k = 
(1-\alpha)\left(\frac{1 + \alpha}{2}\right)^h $$

 \section{Rearrangements  of vertices}\label{sec:reargn}
 
We begin our explorations on the possible modifications on the tree structure
that will improve the valuation of PageRank.
Our first result shows that completely erasing the vertices farthest away
from the root  improves the PageRank. This corroborates the known fact that
the optimal configuration is a {\bf star}, 
i.e. a rooted tree of height $1$ (see e.g. \cite{BGS05}, \cite{LM04}).

 \begin{theorem}\label{fullborra}
If in a  tree $\T^r = 1n_1\ldots n_h$ of height $h \ge 1$, the last level $N_h$ is completely erased, then the PageRank of its root $r$,
 $\PR(h)$, increases its value.
\end{theorem}
\pf
After passing from the tree $\T^r = 1n_1\ldots n_h$, with 
$N = 1 + n_1 + \ldots + n_h$ vertices and PageRank $\PR(h)$, to the tree 
$\T^{'r} = 1n_1\ldots n_{h-1}$ with $N-n_h$ vertices and PageRank $\PR'(h)$, we get
\begin{eqnarray*}
\PR'(h) &-& \PR(h) \
= \frac{(1-\alpha)n_h}{(N-n_h)N}(1+n_1\alpha + \ldots +n_{h-1}\alpha^{h-1}-(N-n_h)\alpha^h)\\
&=& \frac{(1-\alpha)n_h}{(N-n_h)N}
  (1+n_1\alpha+ \ldots +n_{h-1}\alpha^{h-1}-(1+n_1+ \ldots +n_{h-1})\alpha^h)\\
&=& \frac{(1-\alpha)n_h}{(N-n_h)N}
   ((1-\alpha^h)+n_1(\alpha-\alpha^h)+ \ldots +n_{h-1}(\alpha^{h-1} -\alpha^h))>0 
\end{eqnarray*}
because $0< \alpha <1$ and $h \ge 1$. \qed

\begin{remark} 
Thus, in order to improve the PageRank of the root of a tree one can delete as many levels, from   highest to   lowest, as the context permits. Conversely, if a  new level of vertices is added to  a tree, then the PageRank of its root decreases.
\qed
\end{remark}

If it were the case that for practical, or any other  reason, we were obliged to keep certain height, then a natural  question is how much can we prune the tree to improve on PageRank. The extreme situation is to prune all but one arc at each level, so we 
take that structure as 
benchmark and called it  {\em queue tree}.
\begin{definition}\label{qtree}
The {\bf queue tree} of a  tree $\T^r = 1n_1\ldots n_h$ is the tree
$$\T^r_q = 1n_1\ldots n_{\lfloor \frac{h-1}{2}\rfloor}\underbrace{1 \ldots 1}_{\lfloor \frac{h}{2}\rfloor+1}$$
\end{definition}

\begin{theorem}\label{prqtree}
The PageRank of the root of a  tree is smaller than the PageRank of the root of its queue tree.
\end{theorem}
\pf
We proceed recursively from the last level down to $\lfloor \frac{h-1}{2}\rfloor$. 

\noindent
 $(a)$ The PageRank $\PR(h)$ of the root $r$ of $\T^r= 1n_1 \ldots n_{h-1}n_h$ is smaller than the PageRank $\PR'(h)$ of $\T^{'r}= 1n_1 \ldots n_{h-1}1$. Indeed, let 
$N=1+n_1+\ldots +n_h$, then
\begin{eqnarray*}
\PR'(h)&-&\PR(h) = \frac{(n_h-1)(1-\alpha)}{(N-(n_h-1))N}\left(
 \sum_{k=0}^{h-1}n_k\alpha^k - (N - n_h)\alpha^h \right)\\
   &=& \frac{(n_h-1)(1-\alpha)}{(N-(n_h-1))N}
 \sum_{k=0}^{h-1}n_k(\alpha^k - \alpha^h)  >0
\end{eqnarray*}
Apply the same methodology for $\T^r= 1n_1$ \ldots $n_{h-2}n_{h-1}1$ and
$\T^{'r}= 1n_1$ \ldots $n_{h-2}11$, and so on, up to $\lfloor h/2\rfloor$. At this last step we have\\
$(b)$ $\T^r= 1n_1 \ldots n_{\lfloor\frac{h-1}{2}\rfloor}n_{\lfloor\frac{h+1}{2}\rfloor}
\underbrace{1 \ldots 1}_{\lfloor\frac{h}{2}\rfloor}$, and we shall see that its PageRank is less than that of the queue tree 
$\T^{'r}= 1n_1 \ldots n_{\lfloor\frac{h-1}{2}\rfloor}%
\underbrace{1 \ldots 1}_{\lfloor\frac{h}{2}\rfloor + 1}$. We work   separately  the cases of $h$ even or $h$ odd. 

\noindent
$(b.i)$ If $h=2p-1$ then $\T^r= 1n_1 \ldots n_{p-1}n_{p}
\underbrace{1 \ldots 1}_{p-1}$, $\T^{'r}= 1n_1 \ldots n_{p-1}
\underbrace{1 \ldots 1}_{p}$ and  $N=n_1+\ldots +n_p+p$. Let 
$M=\frac{(n_{p}-1)(1-\alpha)}{(N-(n_{p}-1))N}$. Then

\begin{eqnarray*}
\PR'(h) & - & \PR(h) = M\left(1+ \sum_{k = 1}^{p-1}n_k\alpha^k - (N-n_{p})\alpha^{p} +\sum_{k = p+1}^{2p-1}\alpha^k\right) \\
 &=& M\left((1 - \alpha^p) + \sum_{k = 1}^{p-1}n_k(\alpha^k - \alpha^{p}) +\sum_{k = p+1}^{2p-1}(\alpha^k - \alpha^p)\right) \\
  &=& M\left((1 - \alpha^p) + \sum_{k = 1}^{p-1}(n_k - \alpha^{p-k})(\alpha^k - \alpha^p) \right) >0 
\end{eqnarray*}
\noindent
$(b.ii)$ If $h=2p$ then $\T^r= 1n_1 \ldots n_{p-1}n_{p}
\underbrace{1 \ldots 1}_{p}$, $\T^{'r}= 1n_1 \ldots n_{p-1}
\underbrace{1 \ldots 1}_{p+1}$ and  $N=n_1+\ldots +n_p+p+1$.
One then shows $\PR'(h) - \PR(h) > 0$ by a similar argument as in 
$(b.i)$.
\qed

\begin{remark} 
Theorem \ref{prqtree} can not be improved, in the sense that deleting further vertices (but keeping the height) in a queue tree may or may not improve the PageRank of the root. For small values of $h$, the queue tree is the optimal pruning of a tree for increasing PageRank. 
For example, if $h = 4$ the corresponding queue tree is 
$\T^r_q = 1n_1111$ with PageRank $\PR(h)$, and if $n_1 > 1$ and we remove a vertex from  level $N_1$, we get the tree $\T^{'r} = 1(n_1 - 1)111$ with PageRank $\PR'(h)$, and their difference is
$$
\PR'(h) - \PR(h) = \frac{1-\alpha}{(n_1+3)(n_1+4)}%
(1-4\alpha + \alpha^2 + \alpha^3 +\alpha^4) < 0
$$
for any  $\alpha$ such that  $0.27568 < \alpha < 1$.

For larger values of $h$, an improvement of PageRank will depend on $\alpha$ and on the cardinalities of the levels $N_1$, \ldots, 
$N_{\lfloor \frac{h-1}{2}\rfloor}$.  There are also some improvements that can be done on queue trees of particular trees, such as $m$--ary and binomial. 
\qed
\end{remark}

\section{Hierarchies of trees by height and size}
In what follows we assume that $\frac{1}{2} < \alpha < 1$, an interval of useful values for $\alpha$ in practice (see the analysis on this subject in \cite{LM04}).
We want to order the $m$--ary and binomial trees with respect to their PageRank. Which tree structure is best for PageRank? 
 Our first result on this theme gives a hierarchy with respect to the height.
\begin{theorem}\label{orderTheight}
For values of the height $h$ sufficiently large, we have
$$\PR_{1}(h) > \PR_{b}(h)  > \PR_{2}(h)  > \PR_{3}(h)  > \ldots > 
\PR_{m}(h)  \ldots$$
\end{theorem}
\pf  
We have to compute  the appropriate limits: 
 \begin{enumerate}
 \item
 For $1 < k < m$, 
 $\displaystyle \lim_{h \to \infty} \frac{\PR_{k}(h)}{\PR_{m}(h)} = 
 \frac{(k-1)(m\alpha -1)}{(m-1)(k\alpha -1)} > 1$,
 from where we conclude that $\PR_{k}(h) > \PR_{m}(h)$.
 \item
 $\displaystyle    \frac{\PR_{2}(h)}{\PR_{b}(h)} = 
 \frac{\alpha + 1}{2(2\alpha -1)}\frac{(2\alpha )^{h+1}-1}{(\alpha +1)^{h+1}}
\frac{2^{h+1}}{2^{h+1}-1} \stackrel{h\rightarrow \infty }{\longrightarrow } 0$ .
 \item
 $\displaystyle    \frac{\PR_{b}(h)}{\PR_{1}(h)} = \frac{(1-\alpha )(h+1)
\left(\frac{1+\alpha }{2}\right)^h}{1-\alpha ^{h+1}}
\stackrel{h\rightarrow \infty }{\longrightarrow }  0$ .
 \qed
 \end{enumerate}

Next we classify the PageRank of the $m$--ary and binomial trees with respect to their order. (Beware that we will express the order in terms of the height, and therefore we keep the notation   $\PR_{m}(h)$ and $\PR_{b}(h)$ that remarks the dependency of PageRank on  the height of the tree.)

\begin{theorem}\label{Tsize}
For sufficiently large order $N$ and $\alpha \ge  0.58$, we have
$$\ldots \PR_{m}(h) > \ldots   > \PR_{5}(h)  >  \PR_{b}(h)  > \PR_{4}(h)  > \PR_{3}(h)  > \PR_{2}(h)  > \PR_{1}(h)  $$
\end{theorem}
\pf 
The proof splits into three cases. 

\noindent
$(a)$ {\em If $1< k < m$ and $N >> 0$ then $\PR_{m}(h) > \PR_{k}(h)$ }: 
A $k$--ary tree of height $h$ has $\displaystyle N = \frac{k^{h+1}-1}{k-1}$ many vertices. An $m$--ary tree of height $h'$ has the same number of vertices as in a $k$--ary tree if, and only if, 
$\displaystyle \frac{k^{h+1}-1}{k-1}=\displaystyle \frac{m^{h'+1}-1}{m-1}$, 
or equivalently
$h'+1=log_m\left(\frac{m-1}{k-1}(k^{h+1}-1)+1\right)\approx 
log_m\left(\frac{m-1}{k-1}(k^{h+1})\right)$, where the last relation indicates an equivalence among infinite large quantities. Then
\begin{eqnarray*}
\lim_{h\to \infty} \frac{\PR_{k}(h)}{\PR_{m}(h)}  &=& 
\lim_{h\to \infty} \frac{(m\alpha -1)}{(k\alpha -1)}\frac{(k\alpha )^{h+1}-1}{(m\alpha )^{h'+1}-1} \\
&=&
\lim_{h\to \infty} \frac{(m\alpha -1)}{(k\alpha -1)}\frac{1}{(m\alpha )^{log_m}\frac{m-1}{k-1}}
\left(\frac{\alpha }{\alpha ^{log_mk}}\right) ^{h+1}
= 0
\end{eqnarray*}
since $ k<m$ implies $log_mk<1$, and in consequence
$\displaystyle\frac{\alpha }{\alpha ^{log_mk}} <1.$

\noindent
$(b)$  {\em If $m \ge 2$ and $N >> 0$ then $\PR_{m}(h) > \PR_{1}(h)$}:

$$  \displaystyle 
\frac{\PR_{1}(h)}{\PR_{m}(h)}=
\frac{ \frac{1-\alpha^{h'+1}}{h' +1}}
{{\frac{(m-1)(1-\alpha )}{m^{h+1}-1} 
 \frac{(m\alpha )^{h+1}-1}{m\alpha -1}}}
=  \frac{(m\alpha -1)}{(1-\alpha) } 
\frac{(1-\alpha^{\frac{m^{h+1}-1}{m-1}})}{(m\alpha )^{h+1}-1}
\stackrel{h\rightarrow \infty }{\longrightarrow } 0
$$

\noindent
$(c)$  {\em If $N >> 0$ then $\PR_{5}(h) > \PR_{b}(h) > \PR_{4}(h)$}:
A binomial tree of height $h'$ has $2^{h'}$ vertices. For $m\ge 2$, an $m$--ary tree of height $h$ has same cardinality of a binomial tree of height $h'$ if, and only if,   $\displaystyle \frac{m^{h+1} -1}{m-1}=2^{h'}$, or 
equivalently, 
$$\displaystyle 
h'=\log_{_2}\frac{m^{h+1}-1}{m-1}\approx \log_{_2}\frac{m^{h+1}}{m-1}.
$$
 We then show that
 $$ \lim_{h \to \infty} \frac{\PR_{m}(h)}{\PR_{b}(h)}=
 \frac{(1+\alpha)^{\log_{_2}(m-1)}}{m\alpha -1}\left(\frac{m\alpha }
{(1+\alpha )^{\log_{_2}m}}\right)^{h+1}=L $$ 
The limit $L$ is $0$ or $\infty$ depending on $\frac{m\alpha }
{(1+\alpha )^{\log_{_2}m}}$ being less than or greater than 1, respectively.
We showed $L=0$ if $m\le 4$ or $m = 5$ and $\alpha < \alpha_0$, where $\alpha_0$ is the irrational number solution
of $5\alpha_0 = (1 + \alpha_0)^{\log_{_2}5}$, namely, $\alpha_0 = 0.57016 \ldots$.
On the other hand,  $L= \infty$ if $m\ge 6$ or $m = 5$ and
$\alpha > \alpha_0$.
\qed 
  
\subsection{Refining the hierarchy}
According to our results there are many non uniform ways in which one can improve the PageRank in our binomial or $m$--ary trees, yet keeping the height as a constraint for maintaining a hierarchically organised website: it is sufficient to remove vertices at farther distance from the root. However, there are also non trivial uniform trees with better PageRank than the binomial, with respect to height (Theorem \ref{orderTheight}). We present one such possibility which can be seen as a basic backbone  for a website  with different levels.
Our intention with this example is to illustrate the following fact:  to optimise the PageRank values of certain pages of a web site is in general a hard task, as there could be exponentially many possible rearrangements. We will come back to this point in section \ref{sec:tips4uniT}.

We define the 
{\bf path tree of height}  $h$, denoted $\T^r_{\rho}(h)$, as a tree with a root from which hangs $h$ paths   of $h$,
$h-1$, \ldots, $2$, and $1$ vertices.
Observe that $\T^r_{\rho}(h)$ has $h$ vertices at level 1 (connected to the root), $h-1$ vertices at level 2, $h-2$ vertices at level 3, \ldots, one vertex at level $h$. Hence, $|\T^r_{\rho}(h)| = 1 + h(h+1)/2$ and if $r$ is the root of $\T^r_{\rho}(h)$, we have its PageRank, $\PR_{\rho}(h)$, is given by

\begin{eqnarray*}
\PR_{\rho}(h) &=&  \frac{2(1-\alpha) }{2+(h+1)h} 
(1 + h\alpha + (h-1)\alpha^2 + (h-2)\alpha^3 + \ldots + 2\alpha^{h-1} 
+ \alpha^h) \\
&=& \frac{2(1-\alpha) }{2+(h+1)h}
\left( 1 + \alpha^{h+1}\sum_{k=1}^h k \left(\frac{1}{\alpha}\right)^k \right)\\
&=& \frac{2(1-\alpha) }{2+(h+1)h}  \left(1+
\frac{\alpha^{h+2}}{(1-\alpha )^2}
  \left(1 -(h+1)\left(\frac{1}{\alpha}\right)^{h} +
  h\left(\frac{1}{\alpha}\right)^{h+1}\right)\right)
\end{eqnarray*}
We then show that
$$
 \frac{\PR_{\rho}(h)}{\PR_b(h)} = \frac{2}{2+(h+1)h}\left(\frac{2}{1+\alpha}
 \right)^h\left(1 + \frac{\alpha^{h+2} + \alpha(1-\alpha)h - \alpha^2}{(1-\alpha)^2}\right)
   \stackrel{h\rightarrow \infty }{\longrightarrow } \infty
  $$
  Thus $\PR_{\rho}(h) > \PR_{b}(h)$ for sufficiently large $h$, although we have checked the inequality computationally for values of $h \ge 3$
  (for $h <3$ the binomial tree and the chain tree are the same). 
  
  On the other hand, one can show that 
  $\ \displaystyle \frac{\PR_{\rho}(h)}{\PR_1(h)} \stackrel{h\rightarrow \infty }{\longrightarrow }  2\alpha $, and hence the relative position of $\PR_{\rho}(h)$ and
  $\PR_1(h)$ in the hierarchy depends on whether $\alpha$ is $> 1/2$ or $< 1/2$.
  
\subsection{Hierarchies for queue trees}
Surprisingly, for the queue trees of $m$--ary and binomial trees, the same ordering of their PageRank with respect to height and order holds. 
For the $m$--ary (resp. binomial) tree of height $h$, we use 
$\PR_{{\rm q},m}(h)$ (resp. $\PR_{{\rm q},b}(h)$) to denote the PageRank of (the root  of) its queue tree. These values will depend on the parity of the height $h$, since by Definition \ref{qtree},  if $h = 2p-1$ then $\T^r_q = 1n_1 \dots n_{p-1}\underbrace{1 \dots 1}_{p}$, and if $h = 2p$ then 
$\T^r_q = 1n_1 \dots n_{p-1}\underbrace{1 \dots 1}_{p+1}$.
Thus, for $m > 1$,
\begin{eqnarray*}
\PR_{{\rm q},m}(2p-1) &=&  \frac{1-\alpha }
  {\frac{m^p-1}{m-1}+p}
  \left(\frac{(m\alpha )^{p}-1}{m\alpha -1}+\alpha ^p\frac{\alpha ^p-1}{\alpha -1}\right)\\
\PR_{{\rm q},m}(2p)&=&  \frac{1-\alpha }
  {\frac{m^p-1}{m-1}+p+1}
  \left(\frac{(m\alpha )^{p}-1}{m\alpha -1}+\alpha ^p\frac{\alpha ^{p+1}-1}{\alpha -1}\right)
  \end{eqnarray*}
  (the case $m =1$ is trivial since the 1-ary queue tree coincides with the 1-ary tree). And,
  
  \begin{eqnarray*}
\PR_{{\rm q},b}(2p-1)&=&  \frac{1-\alpha }{2^{2p-2}+p}
  \left(\sum\limits_{k=0}^{p-1}{2p-1 \choose k}\alpha ^k+ \alpha ^p\frac{\alpha ^p-1}{\alpha -1}\right)\\
\PR_{{\rm q},b}(2p)&=&  \frac{1-\alpha }
   {2^{2p-1}-\frac{1}{2}{2p \choose p}+p+1}
  \left(\sum\limits_{k=0}^{p-1}{2p  \choose k}\alpha^k+ 
\alpha^p\frac{\alpha^{p+1}-1}{\alpha -1}\right)
\end{eqnarray*}

\begin{theorem}\label{qthierar}
$(i)$ For   sufficiently large height $h$, we have
$$\PR_{{\rm q},1}(h) > \PR_{{\rm q},b}(h) > \PR_{{\rm q},2}(h) > 
\PR_{{\rm q},3}(h) > \ldots > \PR_{{\rm q},m}(h) \ldots$$
$(ii)$ For sufficiently large order $N$ and $\alpha \ge 0.58$, we have
$$\ldots \PR_{{\rm q},m}(h) > \ldots > 
 \PR_{{\rm q},5}(h) >  \PR_{{\rm q},b}(h) > \PR_{{\rm q},4}(h) > \PR_{{\rm q},3}(h) > 
\PR_{{\rm q},2}(h) > \PR_{{\rm q},1}(h) $$
\end{theorem}
The proofs of $(i)$ and $(ii)$ are more involved and longer than previous theorems. We shall give sufficient pointers so that readers may reproduce them.

\medskip
\noindent
{\bf Part $(i)$:} We need to study separately the cases of $h$ being even or odd. The crucial observation is:
\begin{lemma}
 If $0<\alpha<1$ then $$\;(1+\alpha )^{2p-2}\leq \sum\limits_{k=0}^{p-1}{2p-1 \choose k}\alpha ^k + 
 \alpha ^p\frac{\alpha ^p-1}{\alpha -1}\leq (1+\alpha )^{2p-1}$$
\end{lemma}
(To show this observe that for $0< \alpha< 1$ then the quotient of
$\; (1+\alpha )^{2p-1}\; $ and 
$\;\sum\limits_{k=0}^{p-1}{2p-1 \choose k}\alpha ^k + 
 \alpha ^p\frac{\alpha ^p-1}{\alpha -1}\;$ tends to a  constant $L$, 
 with $1 \le L \le 1+\alpha$, as $p$ grows.)
 
Using this lemma, we obtain the same  limits 1), 2) and 3) in Theorem 
\ref{orderTheight} for $h = 2p$ and for $h = 2p-1$.

\medskip
\noindent
{\bf Part $(ii)$:} As in $(i)$ we need to study separately the cases of $h$ being even or odd. 
Observe that for the same type  $\tau$ of queue tree ($\tau$ being 
$m$-ary, binomial, etc.), by Theorem \ref{fullborra}, the queue tree of
height $2p-1$ is obtained by deleting the level $N_{2p}$ of the queue tree
of height $2p$, and hence,
$$\PR_{{\rm q},\tau}(2p-1) > \PR_{{\rm q},\tau}(2p)$$
Therefore, for any two  queue trees of types $\tau_1$ and $\tau_2$
we have
$$\frac{\PR_{{\rm q},\tau_1}(2p-1)}{\PR_{{\rm q},\tau_2}(2p)} >
\max\left\{\frac{\PR_{{\rm q},\tau_1}(2p)}{\PR_{{\rm q},\tau_2}(2p)} ,
\frac{\PR_{{\rm q},\tau_1}(2p-1)}{\PR_{{\rm q},\tau_2}(2p-1)} \right\} >
\frac{\PR_{{\rm q},\tau_1}(2p)}{\PR_{{\rm q},\tau_2}(2p-1)}
$$
and since all these quotients are positive, if any of them reduces to zero 
then all those that are to the right hand side (the ones that are smaller) will also reduce to zero. Hence, to prove that 
$\PR_{{\rm q},\tau_1}(h) < \PR_{{\rm q},\tau_2}(h)$, it will be enough to
prove that 
$\displaystyle 
\frac{\PR_{{\rm q},\tau_1}(2p-1)}{\PR_{{\rm q},\tau_2}(2p)} 
\stackrel{p\rightarrow \infty }{\longrightarrow } 0$.

Now to obtain the inequalities claimed in  $(ii)$ follow the scheme of Theorem 
\ref{Tsize}:
\begin{itemize}
\item[$(a)$] If $1< k<m$ then show 
$\displaystyle 
\frac{\PR_{{\rm q},k}(2p-1)}{\PR_{{\rm q},m}(2p)} 
\stackrel{p\rightarrow \infty }{\longrightarrow } 0$.
\item[$(b)$] If $m > 1$ then show
$\displaystyle 
\frac{\PR_{{\rm q},1}(h)}{\PR_{{\rm q},m}(2p-1)} 
\stackrel{p\rightarrow \infty }{\longrightarrow } 0$,
where  $h = |\PR_{{\rm q},m}(2p-1)|$ (so this $h$ depends on $p$).
\item[$(c)$] To show 
$ \PR_{{\rm q},5}(h) >  \PR_{{\rm q},b}(h) > \PR_{{\rm q},4}(h)$, due to the  observation that $\PR_{{\rm q},\tau}(2p-1) > \PR_{{\rm q},\tau}(2p)$, it will be enough to show that 
$\PR_{{\rm q},5}(2p) >  \PR_{{\rm q},b}(2p-1)$ and
$\PR_{{\rm q},b}(2p) >  \PR_{{\rm q},4}(2p-1)$.
These two inequalities are shown by taking analogous limits as in the proof of part $(c)$ in Theorem \ref{Tsize} and applying the same constraints about $\alpha$.
\end{itemize}

\section{The problem of optimising the link structure}\label{sec:tips4uniT}
As mentioned in the introduction, a theoretically as well as commercially important problem is to find a scheme for modifying the link structure of a local web in order to improve its ranking, as set by PageRank or any other ranking function. 
In this paper we have presented the most fundamental  goal of designing a local web (or fixing an already existing one) with a tree--like structure,  where the PageRank of the main page, located  at the root of the tree, should have the highest possible value,
but at the same time the overall structure of the web should satisfy certain conditions given by the context. We shall not make precise the details of the context, but are the general conditions imposed by design. Let us refer to the context as $\Pi$. 
By virtue of Theorem \ref{pr4treethm} this translates into  the following
optimisation  problem.

\noindent
{\em Main Objective}: Given a certain context $\Pi$, to maximise  the function 
$$\PR(h, 1, n_1, \ldots, n_h) = 
 \frac{1 - \alpha}{1+n_1+\ldots+n_h}  
\sum_{k = 0}^{h} \alpha^{k} n_k$$
for fixed $\alpha$, such that $0< \alpha < 1$, and all trees $\T^r = 1n_1\ldots n_h$ with integer values $h, n_i \ge 1$, $1 \le i \le h$.
If the total number $N$ of vertices is bounded then we can assure that the maximum exists. The complexity of the problem depends mostly on the conditions imposed by the context $\Pi$.
This justifies approaching  the solution through heuristics. Here we give an ad hoc list of rules that clearly stem from our theorems.

\noindent
{\bf\em Rule 1}: Due to  Theorem \ref{fullborra}, the first action  to take is 
to reduce the height as much as the context allows. 

\noindent
{\bf\em Rule 2}: Keep in mind that while applying Rule 1 (and deleting levels),  links between consecutive levels can be rearrange in any  way you like,  as long as the context is kept consistent, and this has no effect on the root's PageRank value (by Theorem \ref{pr4treethm}).

\noindent
{\bf\em Rule 3}: Once the optimal height $h > 1$ is attained\footnote{Optimality  here again depends on maintaining the   context consistent. This height could mean the minimal levels of a hierarchy that we need to 
reflect in the web site; say, for example, of a corporation or a hypertext.}, 
we delete (as much as possible) vertices from levels in the upper half of the tree, trying to get  it close to its underlying queue tree (Theorem \ref{prqtree}), and those vertices that cannot be deleted should be moved    as closer  to level 1 as possible (by Theorem \ref{pr4treethm}).

\medskip
The above rules of general nature can be complemented by next working on the particularities of the queue tree structure. 
For example,  if the applications of rules 1 to 3 give as a final result a 5--ary queue tree, then  Theorem \ref{qthierar}--($i$) tell us that pruning more vertices to convert this tree into a binary queue tree, or binomial queue tree (of same height) improves PageRank. 
The caveat is that we have proved Theorem \ref{qthierar} using
continuous calculus and, therefore, cannot  be applied without doubt for small values of the height. To remedy this  deficiency, we have 
computed $\PR_{{\rm q},b}(h)$ and $\PR_{{\rm q},m}(h)$ for various $m$ and many small integer values of $h$, and concluded the following facts, which strengthen Theorem \ref{qthierar}--($i$):

\begin{enumerate}
\item $\PR_{{\rm q},1}(h)  > \PR_{{\rm q},b}(h)  > \PR_{{\rm q},2}(h)$, for
$h \ge 17$.
\item  $\PR_{{\rm q},b}(h)  > \PR_{{\rm q},1}(h)$, for $2 \le h \le 16$.
\item  $\PR_{{\rm q},b}(h)  > \PR_{{\rm q},2}(h)$, for all $h > 1$.
\item  $\PR_{{\rm q},1}(h)  > \PR_{{\rm q},2}(h)$, for $h \ge 15$.
\item $\PR_{{\rm q},2}(h)  > \PR_{{\rm q},1}(h)$, for $2 \le h \le 14$.
\item $\PR_{{\rm q},2}(h) >  \PR_{{\rm q},3}(h) >   \PR_{{\rm q},4}(h) > \ldots > \PR_{{\rm q},m}(h) \ldots$, for $h \ge 9$.
\item  $\PR_{{\rm q},2}(h) <  \PR_{{\rm q},3}(h) <  \PR_{{\rm q},4}(h) < \ldots < \PR_{{\rm q},m}(h) \ldots$, for $h = 3,4$.
\item Theorem \ref{qthierar}--($i$) is ``almost'' true for $h = 5, 6, 7, 8$ (all but except some arity $m$ from 2 to 6).
\end{enumerate}
Now, depending on the  value of the height of the queue tree obtained by rules 1 to 3, we use the appropriate inequality from the above list to guide our pruning correctly and raise the root's PageRank. For example, if we had  arrived to an $m$--ary queue tree of height 17, and $m \ge 2$, we can delete and move vertices, shaping the tree like a $k$--ary queue tree, for some $k < m$, or like a binomial tree.

\section{The bidirectional case}\label{sec:biTree}
We turn now to    trees with  bidirectional as well as unidirectional arcs.
We use $\B^r$ to denote a tree rooted at $r$ with both unidirectional and bidirectional arcs, where all unidirectional arcs points towards the root $r$.
Formally, a digraph $\B^r= (V,A)$ is a {\bf bidirectional   tree with root} $r$ 
if its set of arcs $A$ can be partitioned in two disjoint  sets $A_1$ and $A_2$ such that:
\begin{itemize}
\item $(V,A_1)$ is a partial tree with root $r$ (the underlying   tree 
of $\B^r$), and
\item if $uv \in A_2$ then $vu \in A_1$, and in this case we say that $v$ is the {\bf origin} of the   {\bf bidirectional arc} $vuv$. (Intuitively think of a bidirectional arc
as a 2-cycle.)
\end{itemize}

\begin{figure}[h]
\setlength{\unitlength}{0.2cm}
\begin{picture}(15.0,11.0)
\put(0,2){\circle*{.4}}
\put(1,0){\circle*{.4}}
\put(1,4){\circle*{.4}}
\put(2,0){\circle*{.4}}
\put(2,2){\circle*{.4}}
\put(5,2){\circle*{.4}}
\put(5,4){\circle*{.4}}
\put(5,6){\circle*{.4}}
\put(8,2){\circle*{.4}}
\put(9,2){\circle*{.4}}
\put(9,4){\circle*{.4}}
\put(9,8){\circle*{.4}}
\put(10,2){\circle*{.4}}
\put(13,6){\circle*{.4}}

\put(7,-2){$\B^r$}

\thinlines  
\put(0,2){\line(1,2){1}}  
\put(1,0){\line(1,2){1}}  
\put(1,4){\line(2,1){8}}  
\put(2,0){\line(0,1){2}}  
\put(2,2){\line(-1,2){1}}  
\qbezier(1,4)(2.35,4.5)(2,2) 
\put(2.05,2.7){\vector(0,-1){0.3}}
\put(5,2){\line(0,4){4}}  
\qbezier(5,6)(9.6,7)(9,4) 
\put(9.05,4.7){\vector(0,-1){0.3}}
\put(8,2){\line(1,2){1}}  
\put(9,2){\line(0,2){2}}  
\put(9,4){\line(-2,1){4}} 
\put(10,2){\line(-1,2){1}}  
\put(13,6){\line(-2,1){4}}  

\put(-0.6,3.7){\normalsize $u'$}
\put(2.4,2){\normalsize $v'$}
\put(4.5,6.7){\normalsize $u$}
\put(8.8,8.7){\normalsize $r$}
\put(9.4,4.2){\normalsize $v$}


\put(15,2){\circle*{.4}}
\put(16,0){\circle*{.4}}
\put(16,4){\circle*{.4}}
\put(17,0){\circle*{.4}}
\put(17,2){\circle*{.4}}
\put(20,2){\circle*{.4}}
\put(20,4){\circle*{.4}}
\put(20,6){\circle*{.4}}
\put(23,2){\circle*{.4}}
\put(24,2){\circle*{.4}}
\put(24,4){\circle*{.4}}
\put(24,8){\circle*{.4}}
\put(25,2){\circle*{.4}}
\put(28,6){\circle*{.4}}

\thinlines  
\put(15,2){\line(1,2){1}}  
\put(16,0){\line(1,2){1}}  
\qbezier(16,4)(17.35,4.5)(17,2) 
\put(17.05,2.7){\vector(0,-1){0.3}}
\put(16,4){\line(2,1){8}}  
\put(17,0){\line(0,1){2}}  
\put(17,2){\line(-1,2){1}}  
\put(20,2){\line(0,4){4}}  
\put(23,2){\line(1,2){1}}  
\put(24,2){\line(0,2){2}}  
\put(24,4){\line(-2,1){4}} 
\put(25,2){\line(-1,2){1}}  
\put(28,6){\line(-2,1){4}}  

\put(14.4,3.7){\normalsize $u'$}
\put(17.4,2){\normalsize $v'$}
\put(19.5,6.7){\normalsize $u$}
\put(23.8,8.7){\normalsize $r$}
\put(24.4,4.2){\normalsize $v$}


\put(23,-2){\circle*{.4}}
\put(24,-4){\circle*{.4}}
\put(24,0){\circle*{.4}}
\put(25,-4){\circle*{.4}}
\put(25,-2){\circle*{.4}}
\put(28,-2){\circle*{.4}}
\put(28,0){\circle*{.4}}
\put(28,2){\circle*{.4}}
\put(31,-2){\circle*{.4}}
\put(32,-2){\circle*{.4}}
\put(32,0){\circle*{.4}}
\put(33,-2){\circle*{.4}}

\thinlines  
\put(23,-2){\line(1,2){1}}  
\put(24,-4){\line(1,2){1}}  
\put(24,0){\line(2,1){4}}  
\qbezier(24,0)(25.35,0.5)(25,-2) 
\put(25.05,-1.3){\vector(0,-1){0.3}}
\put(25,-4){\line(0,1){2}}  
\put(25,-2){\line(-1,2){1}}  
\put(28,-2){\line(0,4){4}}  
\put(28,2){\line(-2,1){4}}  
\put(31,-2){\line(1,2){1}}  
\put(32,-2){\line(0,2){2}}  
\put(32,0){\line(-2,1){4}} 
\put(33,-2){\line(-1,2){1}}  

\put(22.4,-0.3){\normalsize $u'$}
\put(25.4,-2){\normalsize $v'$}
\put(27.5,2.7){\normalsize $u$}
\put(32.4,0.2){\normalsize $v$}

\put(31,-6){\circle*{.4}}
\put(32,-4){\circle*{.4}}
\put(32,-8){\circle*{.4}}
\put(33,-8){\circle*{.4}}
\put(33,-6){\circle*{.4}}
\put(36,-6){\circle*{.4}}
\put(36,-4){\circle*{.4}}
\put(36,-2){\circle*{.4}}
\put(39,-6){\circle*{.4}}
\put(40,-6){\circle*{.4}}
\put(40,-4){\circle*{.4}}
\put(41,-6){\circle*{.4}}

\thinlines  
\put(31,-6){\line(1,2){1}}  
\put(32,-8){\line(1,2){1}}  
\put(32,-4){\line(2,1){4}}  
\qbezier(32,-4)(33.35,-3.5)(33,-6) 
\put(33.05,-5.3){\vector(0,-1){0.3}}
\put(33,-8){\line(0,1){2}}  
\put(33,-6){\line(-1,2){1}}  
\put(36,-6){\line(0,4){4}}  
\put(36,-2){\line(-2,1){4}}  
\put(39,-6){\line(1,2){1}}  
\put(40,-6){\line(0,2){2}}  
\put(40,-4){\line(-2,1){4}} 
\put(41,-6){\line(-1,2){1}}  
\qbezier[10](40,-4) (41.5,-4.75) (43,-5.5)  

\put(30.4,-4.3){\normalsize $u'$}
\put(33.4,-6){\normalsize $v'$}
\put(35.5,-1.3){\normalsize $u$}
\put(40.4,-3.8){\normalsize $v$}


\put(41,2){\circle*{.4}}
\put(42,0){\circle*{.4}}
\put(42,4){\circle*{.4}}
\put(43,0){\circle*{.4}}
\put(43,2){\circle*{.4}}
\put(46,2){\circle*{.4}}
\put(46,4){\circle*{.4}}
\put(46,6){\circle*{.4}}
\put(49,2){\circle*{.4}}
\put(50,2){\circle*{.4}}
\put(50,4){\circle*{.4}}
\put(50,8){\circle*{.4}}
\put(51,2){\circle*{.4}}
\put(54,6){\circle*{.4}}

\thinlines  
\put(41,2){\line(1,2){1}}  
\put(42,0){\line(1,2){1}}  
\put(42,4){\line(2,1){8}}  
\put(43,0){\line(0,1){2}}  
\put(43,2){\line(-1,2){1}}  
\put(46,2){\line(0,4){4}}  
\put(49,2){\line(1,2){1}}  
\put(50,2){\line(0,2){2}}  
\put(50,4){\line(-2,1){4}} 
\put(51,2){\line(-1,2){1}}  
\put(54,6){\line(-2,1){4}}  

\put(40.4,3.7){\normalsize $u'$}
\put(43.4,2){\normalsize $v'$}
\put(45.5,6.7){\normalsize $u$}
\put(49.8,8.7){\normalsize $r$}
\put(50.4,4.2){\normalsize $v$}


\put(49,-2){\circle*{.4}}
\put(50,-4){\circle*{.4}}
\put(50,0){\circle*{.4}}
\put(51,-4){\circle*{.4}}
\put(51,-2){\circle*{.4}}
\put(54,-2){\circle*{.4}}
\put(54,0){\circle*{.4}}
\put(54,2){\circle*{.4}}
\put(57,-2){\circle*{.4}}
\put(58,-2){\circle*{.4}}
\put(58,0){\circle*{.4}}
\put(59,-2){\circle*{.4}}

\thinlines  
\put(49,-2){\line(1,2){1}}  
\put(50,-4){\line(1,2){1}}  
\put(50,0){\line(2,1){4}}  
\put(51,-4){\line(0,1){2}}  
\put(51,-2){\line(-1,2){1}}  
\put(54,-2){\line(0,4){4}}  
\put(54,2){\line(-2,1){4}}  
\put(57,-2){\line(1,2){1}}  
\put(58,-2){\line(0,2){2}}  
\put(58,0){\line(-2,1){4}} 
\put(59,-2){\line(-1,2){1}}  

\put(48.4,-0.3){\normalsize $u'$}
\put(51.4,-2){\normalsize $v'$}
\put(53.5,2.7){\normalsize $u$}
\put(58.4,0.2){\normalsize $v$}

\put(57,-6){\circle*{.4}}
\put(58,-4){\circle*{.4}}
\put(58,-8){\circle*{.4}}
\put(59,-8){\circle*{.4}}
\put(59,-6){\circle*{.4}}
\put(62,-6){\circle*{.4}}
\put(62,-4){\circle*{.4}}
\put(62,-2){\circle*{.4}}
\put(65,-6){\circle*{.4}}
\put(66,-6){\circle*{.4}}
\put(66,-4){\circle*{.4}}
\put(67,-6){\circle*{.4}}

\thinlines  
\put(57,-6){\line(1,2){1}}  
\put(58,-8){\line(1,2){1}}  
\put(58,-4){\line(2,1){4}}  
\put(59,-8){\line(0,1){2}}  
\put(59,-6){\line(-1,2){1}}  
\put(62,-6){\line(0,4){4}}  
\put(62,-2){\line(-2,1){4}}  
\put(65,-6){\line(1,2){1}}  
\put(66,-6){\line(0,2){2}}  
\put(66,-4){\line(-2,1){4}} 
\put(67,-6){\line(-1,2){1}}  
\qbezier[10](66,-4) (67.5,-4.75) (69,-5.5)  

\put(56.4,-4.3){\normalsize $u'$}
\put(59.4,-6){\normalsize $v'$}
\put(61.5,-1.3){\normalsize $u$}
\put(66.4,-3.8){\normalsize $v$}

\put(43,-2){\circle*{.4}}
\put(44,-4){\circle*{.4}}
\put(44,0){\circle*{.4}}
\put(45,-4){\circle*{.4}}
\put(45,-2){\circle*{.4}}
\put(45,-6){\circle*{.4}}
\put(46,-8){\circle*{.4}}
\put(46,-4){\circle*{.4}}
\put(47,-8){\circle*{.4}}
\put(47,-6){\circle*{.4}}

\put(44,0){\line(-1,2){1}}  
\put(43,-2){\line(1,2){1}}  
\put(44,-4){\line(1,2){1}}  
\put(45,-4){\line(0,1){2}}  
\put(45,-2){\line(-1,2){1}}  
\put(46,-4){\line(-1,2){1}}  
\put(45,-6){\line(1,2){1}}  
\put(46,-8){\line(1,2){1}}  
\put(47,-8){\line(0,1){2}}  
\put(47,-6){\line(-1,2){1}}  
\qbezier[10](47,-6) (47.75,-7.5) (48.5,-9)  

\put(44.4,-0.3){\normalsize $u'$}
\put(45.4,-2){\normalsize $v'$}
\put(46.4,-4.3){\normalsize $u'$}
\put(47.4,-6){\normalsize $v'$}

\put(51,-6){\circle*{.4}}
\put(52,-8){\circle*{.4}}
\put(52,-4){\circle*{.4}}
\put(53,-8){\circle*{.4}}
\put(53,-6){\circle*{.4}}
\put(53,-10){\circle*{.4}}
\put(54,-12){\circle*{.4}}
\put(54,-8){\circle*{.4}}
\put(55,-12){\circle*{.4}}
\put(55,-10){\circle*{.4}}

\put(52,-4){\line(-1,2){1}}  
\put(51,-6){\line(1,2){1}}  
\put(52,-8){\line(1,2){1}}  
\put(53,-8){\line(0,1){2}}  
\put(53,-6){\line(-1,2){1}}  
\put(54,-8){\line(-1,2){1}}  
\put(53,-10){\line(1,2){1}}  
\put(54,-12){\line(1,2){1}}  
\put(55,-12){\line(0,1){2}}  
\put(55,-10){\line(-1,2){1}}  
\qbezier[10](55,-10) (55.75,-11.5) (56.5,-13)  

\put(52.4,-4.3){\normalsize $u'$}
\put(53.4,-6){\normalsize $v'$}
\put(54.4,-8.3){\normalsize $u'$}
\put(55.4,-10){\normalsize $v'$}

\put(59,-10){\circle*{.4}}
\put(60,-12){\circle*{.4}}
\put(60,-8){\circle*{.4}}
\put(61,-12){\circle*{.4}}
\put(61,-10){\circle*{.4}}
\put(61,-14){\circle*{.4}}
\put(62,-16){\circle*{.4}}
\put(62,-12){\circle*{.4}}
\put(63,-16){\circle*{.4}}
\put(63,-14){\circle*{.4}}

\put(60,-8){\line(-1,2){1}}  
\put(59,-10){\line(1,2){1}}  
\put(60,-12){\line(1,2){1}}  
\put(61,-12){\line(0,1){2}}  
\put(61,-10){\line(-1,2){1}}  
\put(62,-12){\line(-1,2){1}}  
\put(61,-14){\line(1,2){1}}  
\put(62,-16){\line(1,2){1}}  
\put(63,-16){\line(0,1){2}}  
\put(63,-14){\line(-1,2){1}}  
\qbezier[10](63,-14) (63.75,-15.5) (64.5,-17)  

\put(60.4,-8.3){\normalsize $u'$}
\put(61.4,-10){\normalsize $v'$}
\put(62.4,-12.3){\normalsize $u'$}
\put(63.4,-14){\normalsize $v'$}

\end{picture}\par
\vspace{3cm}
\caption{Bidirectional tree $\B^r$ and its infinite associated tree in two stages.}\label{fig:bitree}
\end{figure}
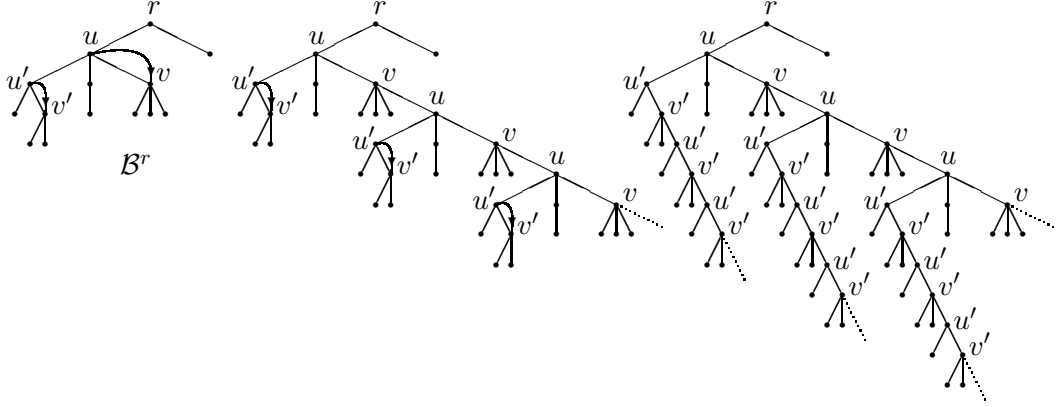

Observe that for each arc $uv \in A_2$ the corresponding bidirectional arc
$vuv$ defines an infinite number of walks ending at the root $r$ (just as would do any cycle within a tree). Henceforth, to the effect of computing the PageRank of $r$ with  
equation (\ref{brinPR}), we can view each arc
$uv \in A_2$ as a path of infinite length hanging from the vertex $v$,
and containing alternatively copies of vertices $u$ and $v$, 
where at each $v$ hangs a copy of the tree rooted at $v$, $\T^v$, and at each $u$ hangs a copy of the remainder of the tree rooted at $u$ after removing from it the sub--tree $\T^v$, that is, $\T^u\setminus \T^v$. Note that $\T^u$ (and $\T^v$) may contain bidirectional arcs.
 Extending this idea through all bidirectional arcs, we can view the bidirectional tree $\B^r$
as an infinite tree. 
 Figure \ref{fig:bitree}  shows  a bidirectional tree $\B^r$ with two bidirectional arcs,
 $vuv$ and $v'u'v'$ (leftmost tree); next to it the bidirectional tree with an infinite branch corresponding to $vuv$; and the rightmost tree is the full infinite tree associated to  $\B^r$.


This view of $\B^r$ as an infinite tree makes it easier to  understand
the interpretations we do below  of equation (\ref{brinPR})
adapted to our trees.
To be clear, what we mean by the 
 infinite tree associated to $\B^r$ is the    tree rooted at $r$, which contains the underlying tree previously defined (i.e. the partial tree rooted at $r$, $(V,A_1)$), and  such that
 for each vertex $v \not= r$ that is the 
origin of a bidirectional arc $vuv$ in $\B^r$, substitute the
arc $uv$ by a countable  infinite   path rooted at $v$, 
 containing alternatively copies of vertices $u$ and $v$, 
 where at each $v$ hangs a copy of the tree rooted at $v$, $\T^v$, and at each $u$ hangs a copy of the remainder of the tree rooted at $u$ after removing from it the sub--tree $\T^v$, that is $\T^u\setminus \T^v$.

Now, 
let us recall equation  (\ref{brinPR}):
$\displaystyle {\PR}(a) = \frac{1-\alpha}{N}\sum_{v \in V}
 \sum_{\rho\,:\,  v\, \stackrel{*}{\longrightarrow}\, a} \alpha^{l(\rho)}D(\rho)$.
 In it, the sum is taken over all vertices $v$ connected through a walk to $a$.
 In an associated infinite tree 
 this walk is a unique path $\rho$ connecting $v$ with $a$. This path could have various incidence  of bidirectional arcs.
 On the other hand, each bidirectional arc $uvu$, with $u\not=r$ and  $od(u) =2$, produces an infinite number of walks: $u$, $uvu$, $uvuvu$, \ldots, with branching factors $D(u) =1$, $D(uvu) = 1/2$, $D(uvuvu) = 1/2^2$, \ldots ; hence, summing over all these walks we get
  $$\sum_{\rho\,:\,  u\, \stackrel{*}{\longrightarrow}\, u} \alpha^{l(\rho)}D(\rho)
 = 1 + \frac{\alpha^2}{2} + \frac{\alpha^4}{2^2} + \cdots = \frac{1}{1-\alpha^2/2} $$
Therefore, if the path  $\rho\,:\,  v\, \stackrel{*}{\longrightarrow}\, a$
contains $q$  vertices, each meeting a bidirectional arc, the contribution to 
$\PR(a)$ of the possible walks produced on $\rho$ is
$\displaystyle \frac{1}{(1-\alpha ^2/2)^q}$.

If the bidirectional arc  is  $rvr$, with $od(r) =1$, and hence $D(rvr\ldots vr) = 1$ for any walk on this arc, we get that the contribution to $\PR(a)$ is
$\displaystyle \frac{1}{(1-\alpha ^2)}$.

All the above observations lead to the following result on computing the PageRank on bidirectional trees.
\begin{theorem}\label{thm:pr4bi}
Let $\B^r = (V,A)$ be a bidirectional tree rooted at $r$.\\
$(1)$ If $od(r) = 0$, then for all $a \in V$, 
\begin{equation}\label{eq:pr4bi:r0}
 \PR(a)=\displaystyle\frac{1-\alpha }{N}\sum_{v \in V}
\frac{\alpha^{l(\rho)}}{2^n(1-\alpha ^2/2)^q}
\end{equation}
$(2)$ If $od(r) = 1$ with bidirectional arc $rur$, then 
\begin{equation}\label{eq:pr4bi:r1a}
\PR(a)=\displaystyle\frac{1-\alpha }{N}
\sum_{v \in V}\frac{\alpha^{l(\rho)}}{2^n(1-\alpha ^2/2)^q}, \; \mbox{ for }\ a\notin \{r,u\}
\end{equation}
and
\begin{equation}\label{eq:pr4bi:r1b}
\PR(a)=\displaystyle\frac{1-\alpha }{N}
\sum_{v \in V}\frac{\alpha^{l(\rho)}}{2^n(1-\alpha ^2/2)^{q-1}(1-\alpha ^2)}, \; \mbox{ for }\  a\in \{r,u\}
\end{equation}
where in all cases, $\rho:\,  v\, \stackrel{*}{\longrightarrow}\, a$ is the unique path from $v$ to $a$, and $l(\rho)$ is the length of this path;
$n$ is the number of bidirectional vertices (i.e. with $od(u) =2$) not being an end-vertex in $\rho$; $q$ is the number of bidirectional arcs meeting $\rho$. \qed
\end{theorem}

In particular, if $od(r)=0$, 
\begin{equation}\label{eq:pr4bi:r0r}
\PR(r)=\displaystyle\frac{1-\alpha }{N}\sum_{v \in V}\frac{\alpha^{l(\rho)}}{(2-\alpha ^2)^q} 
\end{equation}
since $\,n=q$ for this case.
And if $od(r)=1$, 
\begin{equation}\label{eq:pr4bi:r1r}
\PR(r)=\displaystyle\frac{1-\alpha }{N}
\sum_{v \in V}\frac{\alpha^{l(\rho)}}{(2-\alpha ^2)^{q-1}(1-\alpha ^2)} 
\end{equation} 
since $\,n=q-1$ for this case.

\bigskip
At this point we would like to make a digression into the nature of the formulas for PageRank we have just deduced. These have their origin in Brinkmeier's equation (eq. (\ref{brinPR})), which in essence computes the contributions of vertices to the value $\PR(a)$ by a depth-first search exploration. Our proposed equation  for computing the PageRank of the root in the case of unidirectional trees (section \ref{PR4tree}, equation (\ref{pr4tree})) is founded on the complementary tree-search routine, namely, breadth-first search; and we would like to have a result on the same spirit of counting by levels for the case of bidirectional trees.

For a breadth-first search type of computation of PageRank on a bidirectional tree, we must classify  somehow the vertices by levels of the tree. For each $k>0$, the vertices at  level $N_k =\{v_{k1},\ldots, v_{kn_k}\}$ are characterise by the number of bidirectional arcs met by their paths which ends in the root, $v_{ki} \ldots r$.
Hence, $n_k = n_k^0+ \cdots + n_k^{k+1}$, where $n_k^q$ denotes the number of vertices at level $N_k$ having $q$ bidirectional arcs  meeting their paths to $r$. Some of these  $n_k^q$ could be null.
The non-null $n_k^q$ many vertices contributes to the summation in equations
(\ref{eq:pr4bi:r0r}) and (\ref{eq:pr4bi:r1r}) the quantities
$\displaystyle \frac{n_k^q\alpha^k}{(2-\alpha^2)^{q}}$
and
$\displaystyle \frac{n_k^q\alpha^k}{(2-\alpha^2)^{q-1}(1-\alpha ^2)}$
according to either case of $od(r) =0$ or $od(r)=1$. 
Thus, we have the following result.

\begin{theorem}\label{thm:pr4bi:breadth}
Let $\B^r$ be a bidirectional tree rooted at $r$, with $N$ vertices and height $h>0$.\\
$(1)$ If $od(r) = 0$,  
$\qquad \PR(r)=\displaystyle\frac{1-\alpha }{N}\sum_{k =0}^h\sum_{q=0}^{k}
\frac{n_k^q\alpha^{k}}{(2-\alpha ^2)^q}$\\
$(2)$ If $od(r) = 1$, 
$\qquad \PR(r)=\displaystyle\frac{1-\alpha }{N}
\sum_{k =0}^h\sum_{q=0}^{k}\frac{n_k^{q+1}\alpha^{k}}{(2-\alpha ^2)^q(1-\alpha^2)}$\\
where $q$ is the number of bidirectional arcs met by the path ending in $r$, but distinct from the bidirectional arc incidence with $r$, if such bidirectional arc exists. \qed
\end{theorem}

We can give a more succinct vectorial formulation of the previous result, if we develop the sums ``by rows" (outmost sum) and group column terms in a vector.

\begin{theorem}
Let $\B^r$ be a bidirectional tree rooted at $r$ with $N$ vertices and height $h > 0$.
If $od(r) = 0$, then
$\displaystyle
\PR(r) = \frac{1-\alpha}{N}  \sum_{q=0}^h\frac{\Delta_q\cdot \Lambda_q}{(2-\alpha^2)^q}
$,
where $\Delta_q = (n_q^q, n_{q+1}^q, \ldots, n_h^q)$ and
$\Lambda_q = ( \alpha^q, \alpha^{q+1} \ldots, \alpha^h)$.
Similarly, if $od(r) = 1$, then
$\displaystyle
\PR(r) = \frac{1-\alpha}{N}  \sum_{q=0}^h\frac{\Delta'_q\cdot \Lambda_q}{(2-\alpha^2)^q(1-\alpha^2)}
$,
where $\Delta'_q = (n_q^{q+1}, n_{q+1}^{q+1}, \ldots, n_h^{q+1})$. \qed
\end{theorem}

\subsection{Case of $s$-cycles}\label{sec:scycle}
\begin{figure}[h]
\begin{center}
\smallskip
\setlength{\unitlength}{0.7cm}  
\begin{picture}(8,9)          
\thinlines                                     


\put(-0.5,5){\circle* {0.2}}      
\put(0.5,5){\circle* {0.2}}
\put(0,6){\circle* {0.2}}
\put(0,7){\circle* {0.2}}
\put(0,8){\circle* {0.2}}

\put(-0.5,5){\vector(1,2){0.45}}   
\put(0.5,5){\vector(-1,2){0.45}}
\put(0,6){\vector(0,1){0.9}}
\put(0,7){\vector(0,1){0.9}}
\qbezier(0,7)(0.7,6.5)(0,6) 
\put(0.3,6.3){\vector(-1,-1){0.22}}

\put(-0.05,8.25){\small {$r$}}  
\put(0.2,7){\small {$u$}}
\put(0.2,5.9){\small {$v$}}
\put(-0.6,4.6){\small {$s$}}
\put(0.4,4.55){\small {$t$}}

\put(1,6.3){\vector(1,0){1.25}}


\put(2.5,5){\circle* {0.2}}      
\put(3.5,5){\circle* {0.2}}
\put(3,6){\circle* {0.2}}
\put(3,7){\circle* {0.2}}
\put(3,8){\circle* {0.2}}
\put(4.5,6){\circle* {0.2}}
\put(4.5,7){\circle* {0.2}}
	
\put(2.5,5){\vector(1,2){0.45}}   
\put(3.5,5){\vector(-1,2){0.45}}
\put(3,6){\vector(0,1){0.9}}
\put(3,7){\vector(0,1){0.9}}
\put(3,7){\vector(1,0){1.4}}
\put(4.5,7){\vector(0,-1){0.9}}
\put(4.5,6){\vector(-1,0){1.4}}

\put(2.95,8.25){\small {$r$}}  
\put(3.2,7.1){\small {$u$}}
\put(3.25,5.7){\small {$v$}}
\put(4.7,6.95){\small {$v_2$}}
\put(4.7,5.95){\small {$v_1$}}
\put(2.4,4.6){\small {$s$}}
\put(3.4,4.55){\small {$t$}}

\put(5.9,6.3){\normalsize {$=$}}


\put(8,4){\circle* {0.2}}
\put(8,5){\circle* {0.2}}
\put(7.25,5){\circle* {0.2}}      
\put(6.5,5){\circle* {0.2}}
\put(8,6){\circle* {0.2}}	
\put(8,7){\circle* {0.2}}
\put(8,8){\circle* {0.2}}
	
\put(8,4){\vector(0,1){0.9}}   
\put(8,5){\vector(0,1){0.9}}
\put(7.25,5){\vector(3,4){0.7}}
\put(6.5,5){\vector(3,2){1.4}}
\put(8,6){\vector(0,1){0.9}}
\put(8,7){\vector(0,1){0.9}}
\put(4.5,6){\vector(-1,0){1.4}}
\qbezier(8,7)(10,5.5)(8,4) 
\put(8.33,4.28){\vector(-1,-1){0.22}}

\put(7.95,8.25){\small {$r$}}  
\put(8.2,7.){\small {$u$}}
\put(8.2,6.){\small {$v$}}
\put(8.2,5.){\small {$v_1$}}
\put(7.95,3.55){\small {$v_2$}}
\put(6.4,4.6){\small {$s$}}
\put(7.1,4.55){\small {$t$}}


\put(1.5,0){\circle* {0.2}}      
\put(2.5,0){\circle* {0.2}}
\put(2,1){\circle* {0.2}}
\put(2,2){\circle* {0.2}}
\put(2,3){\circle* {0.2}}

\put(1.5,0){\vector(1,2){0.45}}   
\put(2.5,0){\vector(-1,2){0.45}}
\put(2,1){\vector(0,1){0.9}}
\put(2,2){\vector(0,1){0.9}}
\qbezier(2,2)(2.7,1.5)(2,1) 
\put(2.3,1.3){\vector(-1,-1){0.22}}

\put(1.95,3.25){\small {$r$}}  
\put(2.2,2){\small {$u$}}
\put(2.2,0.9){\small {$v$}}
\put(1.4,-0.4){\small {$s$}}
\put(2.4,-0.45){\small {$t$}}

\put(3.5,1.3){\vector(1,0){1.25}}


\put(5.5,0){\circle* {0.2}}      
\put(6.5,0){\circle* {0.2}}
\put(6,1){\circle* {0.2}}
\put(6,2){\circle* {0.2}}
\put(6,3){\circle* {0.2}}

\put(5.5,0){\vector(1,2){0.45}}   
\put(6.5,0){\vector(-1,2){0.45}}
\put(6,1){\vector(0,1){0.9}}
\put(6,2){\vector(0,1){0.9}}
\qbezier(6,2)(7.5,1)(6.5,0) 
\put(6.787,0.3){\vector(-1,-1){0.22}}

\put(5.95,3.25){\small {$r$}}  
\put(6.2,2){\small {$u$}}
\put(6.2,0.9){\small {$v$}}
\put(5.4,-0.4){\small {$s$}}
\put(6.4,-0.45){\small {$t$}}

\end{picture}\par 
\vspace{0.4cm}
\caption{Examples of cyclical   trees.}\label{fig:cyc}
\end{center}
\end{figure}
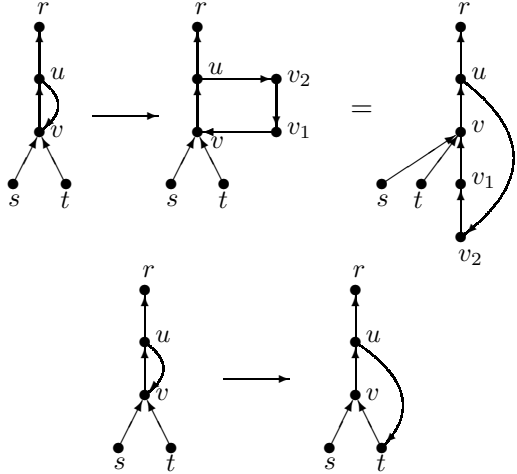

In this section we generalise the computation of PageRank to bidirectional trees 
  of height $h > 1$ on which we close
permissible cycles of any length obtained
by joining vertices from level $N_j$ with vertices from level $N_k$, for $0 \le j < k \le h$. 
In this way we can transform bidirectional arcs $vuv$ into cycles $vuv_n\ldots v_1v$ of longer length, where the arc $uv_n$ close the new cycle inserted in the rooted tree. Also the arc $uv$ of the bidirectional arc $vuv$ can be substituted by
a new arc $ut$ closing a larger path $t\ldots vu$ in the tree. 
In Figure \ref{fig:cyc} we exhibit some examples of these transformations.

Let us call these classes of digraphs obtained by closing cycles on bidirectional trees as {\em cyclical  trees}. Formally 
we define a digraph $\M^r= (V,A)$ as a {\bf cyclical   tree with root $r$},  
if its set of arcs $A$ can be partitioned in two disjoint  sets $A_1$ and $A_2$ such that:
\begin{itemize}
\item $(V,A_1)$ is a partial tree with root $r$ (the underlying   tree of $\M^r$), and
\item if $uv \in A_2$ then there is a path $v_1v_2 \ldots v_{s-1} v_{s}$, beginning at $v_1 = v$, ending at $v_{s}=u$ and with intermediate vertices and 
arcs $v_i v_{i+1}$ in $A_1$, and in this case we say that $v$ is the {\bf origin} of the     cycle $vv_2 \ldots v_{s-1} u v $.
\end{itemize}

We proceed to compute the PageRank of these cyclical trees. 
Similarly to the bidirectional case (which is no other than a 2-cycle), we have that each cycle  $uv\ldots u$ of length $l>2$ 
 and  $od(u) =2$, produces an infinite number of walks: $u$, $uv\ldots u$, $uv\ldots uv\ldots u$, \ldots, with branching factors $D(u) =1$, $D(uv\ldots u) = 1/2$, $D(uv$ \ldots $uv$ \ldots $u) = 1/2^2$, \ldots ; hence, summing over all these walks we get
  $$\sum_{\rho\,:\,  u\, \stackrel{*}{\longrightarrow}\, u} \alpha^{l(\rho)}D(\rho)
 = 1 + \frac{\alpha^l}{2} + \frac{\alpha^{2l}}{2^2} + \cdots = \frac{1}{1-\alpha^l/2} $$
Therefore, if the path  $\rho\,:\,  v\, \stackrel{*}{\longrightarrow}\, a$
contains $q$  vertices, meeting $q$ cycles of length $l_1$, $l_2$, \ldots, $l_q$, respectively, then  the contribution to 
$\PR(a)$ of the possible walks produced on $\rho$ is
$$\displaystyle \frac{1}{1-\alpha ^{l_1}/2} \cdot \frac{1}{1-\alpha ^{l_2}/2}\cdots \frac{1}{1-\alpha ^{l_q}/2}$$

If the cycle   is  $rv\stackrel{l}{\dots }r$, with $od(r) =1$, and hence 
$D(rv\ldots r) = 1$, we get that the contribution to $\PR(a)$ is
$\displaystyle \frac{1}{(1-\alpha ^l)}$.

\begin{theorem}\label{thm:pr4cy}
Let $\M^r = (V,A)$ be a cyclical  tree rooted at $r$.\\
$(1)$ If $od(r) = 0$, then for all $a \in V$, 
\begin{equation*}\label{eq:pr4cy:r0}
 \PR(a)=\displaystyle\frac{1-\alpha }{N}\sum_{v \in V}
\frac{\alpha^{l(\rho)}}{2^n(1-\alpha^{l_1}/2)\cdots (1-\alpha^{l_q}/2)}
\end{equation*}
$(2)$ If $od(r) = 1$ in the cycle $\, rv_1\dots v_{l_{q-1}}r$, then 
\begin{equation*}\label{eq:pr4cy:r1a}
\PR(a)=\displaystyle\frac{1-\alpha }{N}
\sum_{v \in V}\frac{\alpha^{l(\rho)}}{2^n(1-\alpha^{l_1}/2)\cdots (1-\alpha^{l_q}/2)}, \; \mbox{ for }\ a\notin \{r,v_1,\dots ,v_{l_{q-1}}\}
\end{equation*}
and
\begin{equation*}\label{eq:pr4cy:r1b}
\PR(a)=\displaystyle\frac{1-\alpha }{N}
\sum_{v \in V}\frac{\alpha^{l(\rho)}}{2^n(1-\alpha^{l_1}/2)\cdots (1-\alpha^{l_{q-1}}/2)(1-\alpha^{l_q})}, \; \mbox{ for }\ a\in \{r,v_1,\dots ,v_{l_{q-1}}\}
\end{equation*}
where in all cases, $\rho:\,  v\, \stackrel{*}{\longrightarrow}\, a$ is the unique path from $v$ to $a$, and $l(\rho)$ is the length of this path;
$n$ is the number of bidirectional vertices (i.e. with $od(u) =2$) not being an end-vertex in $\rho$; $q$ is the number of cycles meeting $\rho$ and of lengths $\,l_1, l_2, \dots ,l_q$. \qed
\end{theorem}

In particular, if $od(r)=0$, $\,n=q$, and
\begin{equation}\label{eq:pr4bi:rr0r}
\PR(r)=\displaystyle\frac{1-\alpha }{N}\sum_{v \in V}
\frac{\alpha^{l(\rho)}}{(2-\alpha ^{l_1})\dots (2-\alpha ^{l_q})} 
\end{equation}
And if $od(r)=1$, $\,n=q-1$, and
\begin{equation}\label{eq:pr4bi:rr1r}
\PR(r)=\displaystyle\frac{1-\alpha }{N}
\sum_{v \in V}\frac{\alpha^{l(\rho)}}{(2-\alpha ^{l_1})\dots (2-\alpha ^{l_{q-1}})(1-\alpha ^{l_q})} 
\end{equation}

\section{Properties of bidirectional and cyclical trees}\label{sec:properBi}

Analogously  to  the case of unidirectional trees
we shall analyse  in this section the behaviour of PageRank 
on bidirectional, and more general,  cyclical  trees when 
their topology is modified.
Our first result shows that on a unidirectional tree changing unidirectional arcs to bidirectional  enhance the PageRank value of the end-vertices of the transformed arc, but reduces the PageRank of the root of the tree.
 
\begin{theorem}\label{pro1}
 If   in a unidirectional   tree $\T^r$ an arc $vu$, with $u\not= r$,   is changed to a bidirectional arc $uvu$, then $\PR(u)$ and $\PR(v)$ both increase, but $\PR(r)$ decreases. 
\end{theorem}
\pf 
We introduce some notation first. $\PR_x(\T^y)\;$ 
denotes the PageRank of vertex $x$ in the tree $\T^y$ with root $y$;
$n_p(\T^y)$ denotes the number of vertices at level $N_p$ in the tree $\T^y$.
Now, assume that $u$ is at level $N_k$ in the tree $\T^r$ (and, hence, $v \in N_{k+1}$).
Then, we have that
\begin{eqnarray*}
\PR_r(\T^r) &=& \displaystyle\frac{1 - \alpha}{N}\displaystyle\sum_{p = 0}^{h} n_p(\T^r)\alpha^{p} \\
&=& \displaystyle\frac{1 - \alpha}{N}\left(\displaystyle\sum_{p = 0}^{h}  n_p(\T^r-\T^u)\alpha^{p} 
+\sum_{p = k}^{h} n_p(\T^u)\alpha^{p} \right)
\end{eqnarray*}
and, therefore, if $\B^r$ is the bidirectional tree obtained from $\T^r$ by just adding the bidirectional arc $uvu$, we have
$$
\PR_r(\B^r) = \frac{1 - \alpha}{N}\left(\displaystyle\sum_{p = 0}^{h} n_p(\T^r-\T^u)\alpha^{p} 
+\frac{1}{2-\alpha ^2}\sum_{p = k}^{h}n_p(\T^u)\alpha^{p} \right)<\PR_r(\T^r)
$$
which shows that the PageRank of the root $r$ decreases.
On the other hand, the PageRanks of $u$ and $v$ are given by the equations:
\begin{eqnarray*}
\PR_u(\B^u) &=& \displaystyle\frac{1 - \alpha}{N(1 - \alpha^2/2)}\displaystyle\sum_{p = k}^{h}n_p(\T^u)\alpha^{p-k} 
= \frac{1}{1 - \alpha^2/2}\PR_u(\T^r) > \PR_u(\T^r) 
\end{eqnarray*} 
and
\begin{eqnarray*}
\PR_v(\B^v) &=& \frac{1 - \alpha}{N(1 - \alpha^2/2)}\left(\frac{\alpha }{2}\displaystyle\sum_{p = k}^{h}{ n_p(\T^u-\T^v)}\alpha^{p-k}
              + \displaystyle\sum_{p = k+1}^{h} {n_p(\T^v)}\alpha^{p-(k+1)}\right)\\
              &>&\PR_v(\T^v) \qquad \Box
\end{eqnarray*} 

Using same arguments as given for the previous theorem, we can generalized 
the result to the case where the original tree is bidirectional, and some of its unidirectional arc (if any) is promoted to being bidirectional.

\begin{theorem}\label{pro2}
Let $\B^r$ be a bidirectional tree, and let $\B^{'r}$ be the tree resulting from $\B^r$ 
when one of its arcs $vu$, with $u\not= r$ is transformed  into bidirectional 
arc $uvu$. Then
\begin{enumerate}
\item  $\PR_u(\B'^u)= \displaystyle\frac{1}{1 - \alpha^2/2}\PR_u(\B^u)> \PR_u(\B^u)$.  
\item $\PR_v(\B'^u)> \PR_v(\B^u)$.
\item  If $\,u'v'u'\,$ is a  bidirectional arc  intersecting the path $\,uv_1\dots v_k=r,\,$ then \\
$\PR_{u'}(\B'^r)<\PR_{u'}(\B^r)\;$ and $\;\PR_{v'}(\B'^r)<\PR_{v'}(\B^r)$.
\item $\PR_{x}(\B'^r)<\PR_{x}(\B^r)\;$ for all vertex $\,x\,$ in the path $\,v_1\dots v_k=r.$
\item In particular, $\PR_{r}(\B'^r)<\PR_{r}(\B^r).$
\item The vertices which are neither contained in  the path $\,uv_1\dots v_k=r\,$ nor in the bidirectional arcs  intersecting this path
preserve their original  PageRank.  \qed
\end{enumerate}
\end{theorem}

Theorems \ref{pro1} and  \ref{pro2} suggest that in order to increase the PageRank of the root $r$ of a tree we have to directly promote to bidirectional the arcs incidence to $r$. The consequences of this manipulation is summarized in the following theorem, which is a direct consequence of the two previous results.
\begin{theorem}\label{pro3}
Let $\B^r$ be a bidirectional tree, with $od(r) = 0$, and let $\B^{'r}$ be the tree resulting from $\B^r$ 
when one of its arcs $vr$ is transformed  into bidirectional 
arc $rvr$. Then
\begin{enumerate}
\item $\PR_r(\B'^r)= \displaystyle\frac{\PR_r(\B^r)}{1 - \alpha^2}$. 
(Note that for $\alpha = 0.85$ this increment is $\approx 3.6\PR_r(\B^r)$.)  
\item $\PR_v(\B'^r)= \PR_v(\B^r) +\displaystyle\frac{\alpha \PR_r(\B^r)}{1 - \alpha^2}$.
\item $\PR_r(\B'^r)\geq  \PR_v(\B'^r) \Longleftrightarrow \PR_r(\B^r)\geq (1 + \alpha)\PR_v(\B^r)$.
\item All other vertices  (different from $r$ and $v$) preserve their PageRank. \qed
\end{enumerate}
\end{theorem}

For cyclical trees we have results similar to Theorems \ref{pro1}--\ref{pro3}
but factoring out by $1/(1 - \alpha^l)$ in place of 
$1/(1 - \alpha^2)$.

Now, the pruning of the lower levels of a bidirectional tree has mix consequences for the PageRank of the root, as opposed to the positive results obtained for unidirectional trees in section \ref{sec:reargn}.
We illustrate the possible outcomes of pruning lower levels of a bidirectional trees in the figures below. 

In the tree shown in Figure \ref{fig:bit1}, 
 for $\; n\leq 75\;$  and for all $\;m\geq 1$, successive removal of the $m$ vertices of the last level increments the PageRank of the root, $\PR(1)$.
For $n\ge 76$ and for all $m \ge 1$, successive removal of the $m$ vertices of the last level decrements $\PR(1)$.
On the other hand, in the tree shown in Figure \ref{fig:bit2}, 
for $\; n\leq 31\;$  and for all $\;m\geq 1$, successive removal of the $m$ vertices of the last level increments $\PR(1)$.
For $n\ge 32$ and for all $m \ge 1$, successive removal of the $m$ vertices of the last level decrements $\PR(1)$.
\par
\vspace{2.5cm}
\begin{figure}[h]
\begin{center}
\begin{minipage}[t]{5.5cm}
\setlength{\unitlength}{0.35cm}
\begin{picture}(3,4)

\put(0,0){\circle*{.2}}
\put(1,0){\circle*{.2}}
\put(4,0){\circle*{.2}}
\put(2,2){\circle*{.2}}
\put(4,2){\circle*{.2}}
\put(2,4){\circle*{.2}}
\put(5,4){\circle*{.2}}
\put(4,6){\circle*{.2}}
\put(6,8){\circle*{.2}}
\put(6,-2){\circle*{.2}}
\put(7,-2){\circle*{.2}}
\put(10,-2){\circle*{.2}}
\put(8,0){\circle*{.2}}
\put(8,2){\circle*{.2}}
\put(8,4){\circle*{.2}}
\put(8,6){\circle*{.2}}

\put(0,0){\line (1,1){2}}
\put(1,0){\line (1,2){1}}
\put(4,0){\line (-1,1){2}}
\put(2,2){\line (0,1){2}}
\put(4,2){\line (-1,1){2}}
\put(2,4){\line (1,1){4}}
\put(5,4){\line (-1,2){1}}
\qbezier(2,4)(4,3.5)(4,2)
\qbezier(4,6)(5.5,5)(5,4)

\put(6,-2){\line (1,1){2}}
\put(7,-2){\line (1,2){1}}
\put(10,-2){\line (-1,1){2}}
\put(8,0){\line (0,1){6}}
\put(8,6){\line (-1,1){2}}
\qbezier(8,4)(8.8,5)(8,6)

\put(1.2,2){\large $7$}
\put(1.2,4){\large $4$}
\put(3.2,6){\large $2$}
\put(5.7,8.4){\large $1$}
\put(8.3,0){\large $10$}
\put(8.3,2){\large $9$}
\put(8.4,3.8){\large $6$}
\put(8.3,6){\large $3$}
\put(5.3,3.8){\large $5$}
\put(4.3,1.8){\large $8$}
\put(2,-0.9){\large $n$}
\put(8,-2.9){\large $m$}
\put(1.8,0){\large $\dots  $}
\put(7.8,-2){\large $\dots   $}
\end{picture}\par
\vspace{1cm}
\caption{case $od(r)= 0$}\label{fig:bit1}
\end{minipage}
\hspace{1cm}
\begin{minipage}[t]{5cm}

\setlength{\unitlength}{0.35cm}
\begin{picture}(3,3)

\put(0,0){\circle*{.2}}
\put(1,0){\circle*{.2}}
\put(4,0){\circle*{.2}}
\put(2,2){\circle*{.2}}
\put(2,4){\circle*{.2}}
\put(5,7){\circle*{.2}}
\put(6,-2){\circle*{.2}}
\put(7,-2){\circle*{.2}}
\put(10,-2){\circle*{.2}}
\put(8,0){\circle*{.2}}
\put(8,2){\circle*{.2}}
\put(8,4){\circle*{.2}}

\put(0,0){\line (1,1){2}}
\put(1,0){\line (1,2){1}}
\put(4,0){\line (-1,1){2}}
\put(2,2){\line (0,1){2}}
\put(2,4){\line (1,1){3}}
\qbezier(2,2)(2.7,3)(2,4)
\qbezier(5,7)(7,7)(8,4)

\put(6,-2){\line (1,1){2}}
\put(7,-2){\line (1,2){1}}
\put(10,-2){\line (-1,1){2}}
\put(8,0){\line (0,1){4}}
\put(8,4){\line (-1,1){3}}

\put(1.2,2){\large $4$}
\put(1.2,4){\large $2$}
\put(4.2,7){\large $1$}
\put(8.3,0){\large $6$}
\put(8.3,2){\large $5$}
\put(8.3,3.8){\large $3$}
\put(2,-0.8){\large $n$}
\put(8,-2.8){\large $m$}
\put(1.8,0){\large $\dots  $}
\put(7.8,-2){\large $\dots  $}
\end{picture}
\vspace{1cm}
\caption{case $od(r)= 1$}\label{fig:bit2}
\end{minipage}
\end{center}
\end{figure}


The previous results give us some clues on ways of optimising PageRank of tree-like organised sites. 
Obviously these rules for rearrangement
should apply insofar as the context allows.\footnote{We are aware that some of
the rules listed here (and more that could be derived from our results)
 are, to some extend, already in use  by web masters and SEO analysts, but 
 as far as we have seen, without much mathematical justification.}

\begin{description}
\item[{\em Rule 1}]  To augment the PageRank of the root transform incoming arcs bidirectional. Furthermore, link the root with vertices below in the tree (so that  cycles passing by the root are build). 
\item[{\em Rule 2}] To augment the PageRank of a vertex $u$ different from the root, link $u$ with a bidirectional arc to each one of the vertices on the subtree with root $u$ (hence obtaining a cyclical tree). Keep in mind that this enhances the PageRank of $u$ but reduces the PageRank of the root.
One may interpret this action as linking an individual with all its subordinates in a hierarchical organisation.
\end{description}

\section{More complex topologies}

The next natural step is to
upgrade the preceding results on bidirectional and cyclical   trees
to finite  cyclic structures which can be modelled by our infinite trees.
In order to achieve this further extensions we should then visualise an arbitrary 
digraph through its {\em condensation digraph} as the acyclic digraph consisting of its {\em strongly connected components}. 

A digraph $\D = (V,A)$ is {\bf strongly connected}  if for each pair $u$ and $v$ of distinct vertices, there is a 
path joining $u$ with $v$ and a path joining $v$ with $u$. Define $u\equiv v$ provided there are paths joining $u$ with 
$v$ and $v$ with $u$. This is an equivalence relation and, in consequence,
 $V$ is partitioned into equivalence classes 
$V_1, \dots ,V_p $. The $p$ subdigraphs $\D_i = (V_i, A/V_i) $ induced on the sets $V_i$, $i=1,\dots ,p$, are the 
{\bf strong connected components} of $\D$. The digraph $\D$ is strongly connected if and only if it has exactly one 
strong component. The {\bf condensation digraph} of the digraph $\D$ is the acyclic digraph whose vertices are the strong connected components, or SCC, of $\D$, and there is an arc from one SCC $\D_i$ to another $\D_j$, $i\neq j$, if a vertex of $V_i$ is adjacent in $\D$ to a vertex of $V_j$.

Now, the extension of our techniques and procedures to a more general digraph requires that its corresponding condensation digraph be
 a rooted tree  whose SCC behave like the cyclical
 rooted  trees. Not all SCC may have the  required behaviour, but many do so, 
and the key is that each SCC should have a root through which it connects to the
rest of the tree (and by no other vertex), and two such SCC which are adjacent in the
condensation digraph are linked by just one arc in the original digraph.
The SCC which have these properties we shall called {\em PR--digraph}.

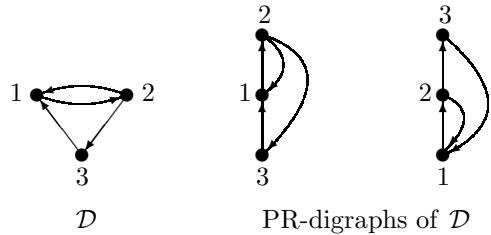
\begin{figure}
\begin{center}
\setlength{\unitlength}{0.8cm}  
\begin{picture}(6,3)          

\thinlines                                     


\put(-0.75,1){\circle* {0.2}}      
\put(0.75,1){\circle* {0.2}}
\put(0,0){\circle* {0.2}}

\put(0,0){\vector(-3,4){0.7}}   
\put(0.75,1){\vector(-3,-4){0.7}}
\qbezier(-0.75,1)(0,1.3)(0.75,1) 
\qbezier(-0.75,1)(0,0.7)(0.75,1)
\put(-0.43,1.11){\vector(-3,-1){0.22}}
\put(0.43,0.90){\vector(3,1){0.22}}

\put(-1.2,0.9){\small {$1$}}  
\put(1.0,0.9){\small {$2$}}
\put(-0.1,-0.5){\small {$3$}}
\put(-0.1,-1.2){\small {$\cal{D}$}}


\put(3,0){\circle* {0.2}}      
\put(3,1){\circle* {0.2}}
\put(3,2){\circle* {0.2}}

\put(3,0){\vector(0,1){0.9}}   
\put(3,1){\vector(0,1){0.9}}
\qbezier(3,2)(3.7,1.5)(3,1) 
\qbezier(3,2)(4.5,1.5)(3,0) 
\put(3.31,1.284){\vector(-1,-1){0.22}}
\put(3.3,0.304){\vector(-1,-1){0.22}}

\put(2.93,2.2){\small {$2$}}  
\put(2.6,0.9){\small {$1$}}
\put(2.9,-0.5){\small {$3$}}
\put(3.0,-1.2){\small {$\mbox{PR-digraphs of}\;\, \cal{D}$}}


\put(6,0){\circle* {0.2}}      
\put(6,1){\circle* {0.2}}
\put(6,2){\circle* {0.2}}

\put(6,0){\vector(0,1){0.9}}   
\put(6,1){\vector(0,1){0.9}}
\qbezier(6,1)(6.7,0.7)(6,0.05) 
\qbezier(6,2)(7.5,0.7)(6,0) 
\put(6.31,0.37){\vector(-1,-1){0.22}}
\put(6.33,0.185){\vector(-3,-2){0.22}}

\put(5.93,2.2){\small {$3$}}  
\put(5.6,0.9){\small {$2$}}
\put(5.9,-0.5){\small {$1$}}

\end{picture}
\vspace{1cm}
\caption{Digraph $\cal D$ and the PR-digraphs that can be derived from it.}\label{fig:digraph}
\end{center}
\end{figure}

Formally, a   {\bf   PR--digraph with root}   $r$ is a strongly connected digraph with 
at least one vertex $r$ (the root of the PR--digraph) such that for all vertex $v$ ($v \not= r$),
there is a unique path joining  $v$ with $r$. In this structure we would say that a 
vertex $v$ is at level $N_k$ if the path that connects $v$ with $r$ is of length $k$.
Now, in essence, a PR--digraph is much like a cyclical    tree in as much as it
can be seen as a tree with cycles formed by adding arcs from one level up to another level down.
The key is that a PR--digraph admits a corresponding infinite tree due to the fact that
each vertex has been assigned a unique level of the graph, or in other words, a unique path
to $r$. Note that the root as the rest of the vertices  may have out--degree greater  or 
equal to 1. 
As an illustration of structures that can be PR--digraphs or not see Figure \ref{fig:digraph}.
There, in the strongly connected digraph $\cal D$ the vertex $1$ can not be the
root of a PR--digraph since vertex $2$ has two  paths towards $1$. 
On the contrary, 
the vertices $2$ and $3$ can be roots of a PR--digraph $\cal D$.

Also the complete digraph $\{(1,2), (2,1), (1,3), (3,1), (2,3), (3,2)\}$ can not be a PR--digraph for any of its vertices. But, the strongly connected digraph $\{(1,2)$,  
$(2,1)$, $(2,3)$, $(3,2)\}$ is a PR--digraph
rooted at any of its three vertices.

The condition characterising a PR--digraph must also apply to the connections
among SCC which are PR--digraphs. It can not be the case that in a tree of SCCs, which
are PR--digraphs, one such SCC connects to another SCC in the tree through two arcs or more;
that is, in the original digraph there must be a root (which itself could be the root of a PR--digraph)
and it must be the case that each vertex $v$ connects to the root by a unique path.
On the other hand, we must admit the possibility of producing cycles of length $s > 1$ in this structure
by connecting a vertex at the level $N_{k+1-s}$ with a vertex at level $N_k$.

\begin{figure}
\begin{center}
\setlength{\unitlength}{0.8cm}  
\begin{picture}(5,4)          

\thinlines                                     

\put(0,1){\circle* {0.2}}      
\put(0.75,0){\circle* {0.2}}
\put(0.75,2){\circle* {0.2}}
\put(1.5,1){\circle* {0.2}}
\put(2.25,3){\circle* {0.2}}
\put(3,0){\circle* {0.2}}
\put(3,1){\circle* {0.2}}
\put(3.75,2){\circle* {0.2}}
\put(4.5,0){\circle* {0.2}}
\put(4.5,1){\circle* {0.2}}

\put(0,1){\vector(3,-4){0.7}}   
\put(0,1){\vector(3,4){0.7}} 
\put(0.75,0){\vector(3,4){0.7}}
\put(0.75,2){\vector(3,-4){0.7}}
\put(0.75,2){\vector(3,2){1.4}}
\put(1.5,1){\vector(-1,0){1.4}}
\qbezier(1.5,1)(1.5,1.95)(0.75,2) 
\put(1.1,1.925){\vector(-3,1){0.22}}
\put(3,0){\vector(0,1){0.9}}   
\put(3,1){\vector(3,4){0.7}} 
\put(3.75,2){\vector(-3,2){1.4}}
\put(3.75,2){\vector(3,-4){0.7}}
\put(4.5,0){\vector(0,1){0.9}}
\put(4.5,1){\vector(-1,0){1.4}}
\qbezier(3,0)(3.75,0.6)(4.5,0) 
\put(3.3,0.18){\vector(-2,-1){0.22}}
\qbezier(3,0)(3.75,-0.6)(4.5,0) 
\put(4.2,-0.18){\vector(2,1){0.22}}

\put(2.2,3.2){\small {$r$}}  
\put(1.7,0.9){\small {$u$}}
\put(2.6,0.9){\small {$y$}}
\put(4.7,0.9){\small {$w$}}
\put(2.6,-0.1){\small {$x$}}
\put(4.7,-0.1){\small {$z$}}
\end{picture}

\vspace{1cm}
\caption{Digraph which is not a PR-digraph tree.}\label{fig:notdi}
\end{center}
\end{figure}
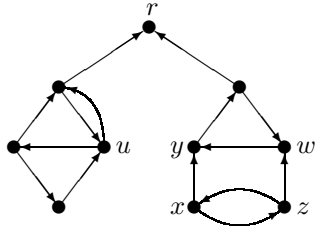

We shall then define a {\bf PR--digraph   tree with root} $r$, as a digraph
${\cal D} =  (V, A)$  whose set of arcs $A$  can be partitioned in two disjoint  
sets $A_1$ and $A_2$ such that:
\begin{itemize}
\item $(V,A_1)$ is a partial digraph whose condensation digraph is a   tree
of  SCCs which are   PR--digraphs, each pair of adjacent PR--digraphs are
linked by a unique arc and the maximal PR--digraph contains the root $r$ 
(the underlying digraph of $\cal D$); and
\item if $uv \in A_2$ then there is a path $v_1v_2 \ldots  v_{s-1}v_s$, beginning at $v = v_1$, ending at $u= v_{s}$ and with intermediate vertices and 
arcs $v_i v_{i+1}$ in $A_1$, and in this case we say that $v$ is the {\bf origin} of the     cycle $ v v_2 \ldots v_{s-1} u v $.
\end{itemize}
Note that this time the arc $uv$, as well as the cycle $v v_2 \ldots v_{s-1}uv$, could be in 
the partial digraph $(V,A_1)$.
We show in Figure \ref{fig:notdi} a digraph $\cal D$ that can not be a PR--digraph   tree. 
This is due to the   fact that the left--side SCC of $\cal D$ is not compatible with a PR--digraph   tree, since the vertex $u$ has out degree 2 towards the root $r$. Also in the SCC on the right branch of $\cal D$ either one of the arcs $xy$ or 
$zw$ represents a surplus  that forbids $\cal D$ from being a PR--digraph   tree. 

\begin{theorem}
A PR--digraph   tree with root $r$ is a cyclical   tree with root $r$.
\end{theorem}
\pf
Let ${\cal D} =  (V, A)$ be a PR--digraph   tree with root $r$. It is sufficient to 
prove that the underlying digraph, ${\cal C} = (V, A_1)$, of $\cal D$ is a cyclical
  tree with root $r$. 
The set of arcs $A_1$ can be partitioned in two disjoint sets:
$A_{11} = \{vu : \mbox{there is a path } vu\ldots r \mbox{ in } {\cal C} \}$
and $A_{12} = A_1 \setminus A_{11}$.
The digraph $(V, A_{11})$ is a tree rooted at $r$ because, by the definition of
the underlying digraph $\cal C$, each vertex of $V$ is joined with the root
$r$ by a unique path. Then  $(V, A_{11})$ is the underlying directed tree of $\cal C$ and, moreover, if $uv$ belongs to $A_{12}$ then $uv$ is in a SCC that is a 
PR--digraph with some root $r'$. By the strong connection, there is a path joining
$v$ to $u$, and thus $\cal C$ is a cyclical   tree.
\qed

\begin{figure}
\begin{center}
\setlength{\unitlength}{0.8cm}  
\begin{picture}(5,5)          

\thinlines                                     

\put(0,2){\circle* {0.2}}      
\put(0.75,1){\circle* {0.2}}
\put(0.75,3){\circle* {0.2}}
\put(1.5,2){\circle* {0.2}}
\put(0.75,4){\circle* {0.2}}

\put(0,2){\vector(3,-4){0.7}}   
\put(0,2){\vector(3,4){0.7}} 
\put(0.75,1){\vector(3,4){0.7}}
\put(0.75,3){\vector(3,-4){0.7}}
\put(0.75,3){\vector(0,1){0.9}}
\put(1.5,2){\vector(-1,0){1.4}}
\qbezier(0.75,4)(1.64,3)(1.5,2) 
\put(1.51,2.4){\vector(0,-1){0.22}}

\put(0.7,4.2){\small {$r$}}  
\put(0.95,2.9){\small {$w$}}
\put(-0.5,1.9){\small {$v$}}
\put(1.75,1.9){\small {$u$}}
\put(0.7,0.5){\small {$t$}}

\put(2.2,2.5){\vector(1,0){1.5}}

\put(4.5,0){\circle* {0.2}}
\put(4.5,1){\circle* {0.2}}
\put(4.55,2){\circle* {0.2}}
\put(4.5,3){\circle* {0.2}}
\put(4.5,4){\circle* {0.2}}

\put(4.5,0){\vector(0,1){0.9}}   
\put(4.5,1){\vector(0,1){0.9}} 
\put(4.5,2){\vector(0,1){0.9}}
\put(4.5,3){\vector(0,1){0.9}}
\qbezier(4.5,4)(6.5,2.0)(4.5,1) 
\put(4.8,1.35){\vector(-1,-1){0.22}}
\qbezier(4.5,3)(5.5,2.0)(4.45,1) 
\put(4.8,1.16){\vector(-2,-1){0.22}}
\qbezier(4.5,2)(3.5,1.0)(4.5,0) 
\put(4.2,0.30){\vector(1,-1){0.22}}

\put(4.45,4.2){\small {$r$}}  
\put(4.7,2.9){\small {$w$}}
\put(4,1.9){\small {$v$}}
\put(4.75,0.9){\small {$u$}}
\put(4.45,-0.5){\small {$t$}}

\end{picture}
\medskip
\caption{PR-digraph tree and associated cyclical tree.}\label{fig:prdi}
\end{center}
\end{figure}
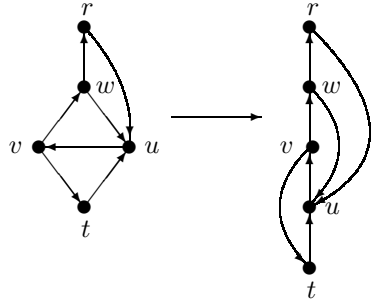

As a consequence of this theorem we can compute the PageRank of the root $r$ of a PR--digraph   tree by a similar formula as given 
  in   section \ref{sec:scycle} for cyclical   trees.
In Figure \ref{fig:prdi} we exhibit a PR--digraph   tree $\cal D$ and its representation as a cyclical   tree.

The PR--digraph   trees are the most general cyclical structures 
which can be interpreted as unidirectional  
 infinite trees, and on which we can apply the optimisation techniques
displayed in this article
by treating each SCC as  one unit. 
This could also revert on a speed up on the PageRank calculation.
More explicitly, the last point we want to call attention to is the following.
There are several approaches in the literature to the task of speeding up the 
calculation of PageRank, based upon the following general scheme (see, for example, \cite{Kamvar, Arasu02, Brink05}): 
\begin{quote}
Partition  the web into local subwebs; then compute some independent ranking for each local subweb, which will apply to the whole subweb treated as a unit; and then compute the ranking of the graph of subwebs.
\end{quote} 

In \cite{Arasu02} and \cite{Brink05} the local splitting of the web is done in strongly connected components, and further in \cite[Thm 2.1]{Brink05}, it is shown that the 
PageRank can be calculated independently on each SCC,    
provided we know the PageRank of all vertices outside the SCC, but  directly linking to vertices in the SCC.
Our PR--digraph   tree is the most simple splitting of the web in the way of 
\cite{Arasu02} and \cite{Brink05}, namely as SCC, with the additional strongest condition of having a single link between components, which by the previously   mentioned result of \cite{Brink05} can have PageRank computed independently on each SCC, and on a very simple way, 
provided we know the PageRank of their descendants in the topological structure of the tree.
This suggests computing PageRank in parallel and through {\em layers},
as it is proposed in  \cite[\S 3]{Brink05},
following an iterated process on the tree from a top level  $N_h$ down to the root at $N_0$.
The PR--digraph is a suitable structure for the application of this process.


\begin{thebibliography}{99}

\bibitem{AJB99} R. Albert, H. Jeong and A.-L. Barabasi. Diameter of the World Wide Web, {\em Nature} 401:130-131, Sep 1999.



\bibitem{Arasu02} A. Arasu, J. Novak, A. Tomkins and J. Tomlin, 
PageRank computation and the structure of the Web: Experiments and algorithms. {\em The Eleventh International World Wide Web Conference, Posters.} 2002.
\bibitem{BGS05} M. Bianchini, M. Gori and F. Scarselli. Inside PageRank, {\em ACM Transactions on Internet Technologies } {\bf 4} (4),  2005.


\bibitem{BP98}  S. Brin and L. Page, The
anatomy of a large scale hypertextual web search engine. {\em Computer
Networks and ISDN Systems}, {\bf 33} (1998), 107-117.

\bibitem{BMPW98}  S. Brin, R. Motwami, L. Page and
Terry Winograd, The PageRank citation ranking: Bringing order to the web. {\em Technical Report}, Comp. Sci. Dept., Stanford University, 1998.

\bibitem{Brink05} M. Brinkmeier.
Distributed calculation of PageRank using strongly connected components,
{\em Proceedings of the I2CS'05 in Paris} (LNCS), 2005.

\bibitem{Brink06} M. Brinkmeier. 
PageRank revisited,  
 {\em ACM Transactions on Internet Technologies}, {\bf  6} (3), 2006.

\bibitem
{CLR}  T. Cormen, C. Leiserson, R. Rivest and C. Stein. {\em Introduction to Algorithms},
The MIT Press, 2nd. Edition, 2001.

\bibitem{florida} P. Craven, Google's ``Florida" Update. In: {\em WebWorks} (http://www.webworkshop.net/florida-update.html), Dec. 3, 2003.

\bibitem{googlepress} Google Acquires Kaltix Corp. In {\em  Google Press} (http://www.google.com/intl/ro/press/pressrel/kaltix.html), 
Sept. 30, 2003.

\bibitem{Kamvar} S. D. Kamvar, T. H. Haveliwala,  C. D. Manning, and G. H. Golub. Exploiting the block structure of the Web for computing PageRank. {\em Stanford University Technical Report}, 2003.


\bibitem{LM04} A. N. Langville and C. D. Meyer, Deeper Inside PageRank,  {\em Internet Mathematics} {\bf 1} (3): 335--380, 2005. 
\bibitem{Olsen03} S. Olsen, Searching for the personal touch. In: {\em CNET News.com} 
(http://news.com.com/2100-1024-5061873.html),  August 11, 2003.

\end{thebibliography}
\end{document}